\documentclass[aps,twocolumn,superscriptaddress,prx]{revtex4-2}

\usepackage{graphicx}
\usepackage{amssymb,amsmath,amsfonts,bm}
\usepackage{subfigure}
\usepackage[normalem]{ulem}
\usepackage{hyperref}
\usepackage[dvipsnames]{xcolor}
\hypersetup{urlcolor=RoyalBlue, colorlinks=true,citecolor=RoyalBlue,linkcolor=RoyalBlue} 
\usepackage{physics}
\usepackage[final]{showkeys}
\usepackage{float}

\newcommand{\hdd}{\includegraphics[scale=0.09,trim={0 10 0 0}]{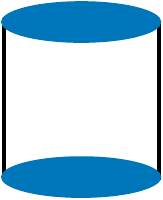}}
\newcommand{\vdd}{\includegraphics[scale=0.09,trim={0 0 0 0}]{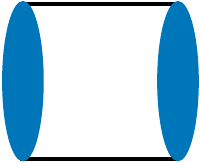}}
\newcommand{\hll}{\includegraphics[scale=0.09,trim={0 0 0 0}]{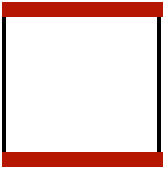}}
\newcommand{\vll}{\includegraphics[scale=0.09,trim={0 0 0 0}]{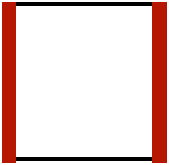}}
\newcommand{\sql}{\includegraphics[scale=0.09,trim={0 0 0 0}]{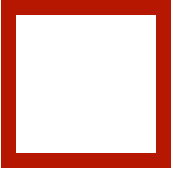}}
\newcommand{\hld}{\includegraphics[scale=0.09,trim={0 0 0 0}]{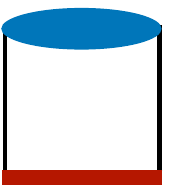}}
\newcommand{\hdl}{\includegraphics[scale=0.09,trim={0 10 0 0}]{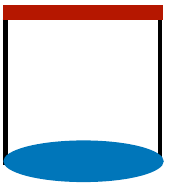}}
\newcommand{\vdl}{\includegraphics[scale=0.09,trim={0 0 0 0}]{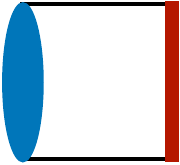}}
\newcommand{\vld}{\includegraphics[scale=0.09,trim={0 0 0 0}]{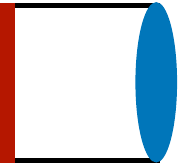}}
\newcommand{\leftopen}{\includegraphics[scale=0.09,trim={0 0 0 0}]{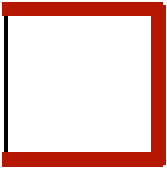}}
\newcommand{\bottomopen}{\includegraphics[scale=0.09,trim={0 0 0 0}]{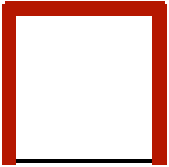}}
\newcommand{\rightopen}{\includegraphics[scale=0.09,trim={0 0 0 0}]{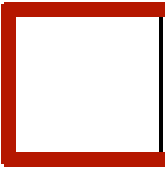}}
\newcommand{\topopen}{\includegraphics[scale=0.09,trim={0 0 0 0}]{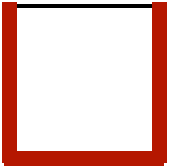}}

\begin{document}
\title{Flux fractionalization transition in anisotropic $S=1$ antiferromagnets and dimer-loop models}
\author{Souvik Kundu}
\affiliation{\small{Tata Institute of Fundamental Research, 1 Homi Bhabha Road, Mumbai 400005, India}}
\author{Kedar Damle}
\affiliation{\small{Tata Institute of Fundamental Research, 1 Homi Bhabha Road, Mumbai 400005, India}}

\begin{abstract}
We demonstrate that the low temperature ($T$) properties of a class of anisotropic spin $S=1$ kagome (planar pyrochlore) antiferromagnets on a field-induced $\frac{1}{3}$-magnetization ($\frac{1}{2}$-magnetization) plateau are described by a model of fully-packed dimers and loops on the honeycomb (square) lattice, with a temperature-dependent relative fugacity $w(T)$ for the dimers. The fully-packed O(1) loop model ($w=0$) and the fully-packed dimer model ($w=\infty$) limits of this dimer-loop model are found to be separated by a phase transition at a finite and nonzero critical fugacity $w_c$, with interesting consequences for the spin correlations of the frustrated magnet. The $w>w_c$ phase has short loops and spin correlations dominated by power-law columnar order (with subdominant dipolar correlations), while the $w<w_c$ phase has dominant dipolar spin correlations and long loops governed by a power-law distribution of loop sizes. Away from $w_c$, both phases are described by a long-wavelength Gaussian effective action for a scalar height field that represents the coarse-grained electrostatic potential of fluctuating dipoles. The destruction of power-law columnar spin order below $w_c$ is driven by an unusual {\em flux fractionalization} mechanism, topological in character but quite distinct from the usual Kosterlitz-Thouless mechanism for such transitions: Fractional electric fluxes which are bound into integer values for $w>w_c$, proliferate in the $w<w_c$ phase and destroy power-law columnar order.
\end{abstract}

\maketitle

\section{Introduction}

Insulating magnets in which the exchange couplings compete due to the geometry of the lattice can display interesting low temperature and low frequency behaviors~\cite{Mila_Mendels_Lacroix_2011book,Moessner_Moore_2021book,Balents_Savary_SpinLiquidsreview2016}. These arise from a large (near) degeneracy of low energy states~\cite{Mila_Mendels_Lacroix_2011book,Moessner_Moore_2021book,Balents_Savary_SpinLiquidsreview2016}. In some particularly interesting cases, the corresponding dynamics is best described in terms of emergent degrees of freedom which provide the natural language for parameterizing the ground state and its elementary excitations~\cite{Moessner_Moore_2021book}. A well-known example of this is the physics of spin-ice materials~\cite{Harris_etal_1997,Siddharthan_Shastry_etal_1999,Fennel_etalScience2009,Castelnovo_Moessner_Sondhi_review2012,Bramwell_Harris_review2020}, best described in terms of fluctuating non-local loop degrees of freedom and emergent electromagnetism~\cite{Jaubert_Haque_Moessner_2011,Huse_Krauth_Moessner_Sondhi_2003,Hermele_Balents_Fisher_2004,Castelnovo_Moessner_Sondhi_2008,Banerjee_Isakov_Damle_Kim_2008,Jaubert_Holdsworth_2009,Ross_Savary_Gaulin_Balents_2011,Lee_Onoda_Balents_2012,Benton_Sikora_Shannon_2012,Henley_Coulombphasesreview2010}. Another well-studied example is the Kitaev model~\cite{Kitaev_annals} and related candidate materials~\cite{Takagi_etal_Kitaevreview2018,Hermanns_Kimchi_Knolle_Kitaevreview2018}.

In this article, we identify an interesting low temperature ($T$) regime in the physics of a class of spin $S=1$ kagome and planar pyrochlore  antiferromagnets with competing strong single-ion anisotropy and exchange anisotropy. We demonstrate that such kagome (planar pyrochlore) antiferromagnets have a field-induced $\frac{1}{3}$-magnetization ($\frac{1}{2}$-magnetization) plateau on which the low temperature behavior is best described by the physics of fluctuating dimer and loop degrees of freedom with a full-packing constraint on the honeycomb (square) lattice. This low-energy theory has a single temperature-dependent coupling $w(T)$ that corresponds to the relative fugacity of dimers. 
This  analysis also relates a class of anisotropic spin $S=1$ models on the pyrochlore lattice to a similar dimer-loop model on the diamond lattice. 

Here, we focus here on the detailed analysis of such kagome and planar pyrochlore systems, leaving aside for now the pyrochlore model, whose low-energy properties require a separate discussion. 
We find that the fully-packed O(1) loop model ($w=0$) and the fully-packed dimer model ($w=\infty$) limits of the corresponding two-dimensional dimer-loop models  are separated by a phase transition at a finite and nonzero critical fugacity $w_c$, with interesting consequences for the spin correlations of the frustrated magnet when it is driven across this transition by varying the temperature.  The $w>w_c$ phase has short loops and spin correlations dominated by power-law columnar order (with subdominant dipolar correlations), while the $w<w_c$ has long loops with a power-law distribution of loop sizes and dominant dipolar spin correlations. We show that this dimer-loop model admits for all $w$ a microscopic description in terms of divergence-free electric polarization field on links of the lattice (equivalently, configurations of a height field $H$ defined on dual lattice sites), 
Away from $w_c$, both phases can be described by a long-wavelength Gaussian effective action for a scalar height field $h$ that represents the coarse-grained electrostatic potential of fluctuating dipoles.

Fractional electric fluxes, which are bound into integer values for $w>w_c$, proliferate in the $w<w_c$ phase and destroy power-law columnar order. Equivalently, in field-theoretical language~\cite{DiFrancesco_Mathieu_Senechal_book_1997}, the transition at $w_c$ corresponds to a spontaneous jump in the ``compactification radius'' of the free scalar field  $h$. Thus,
the destruction of power-law columnar spin order below $w_c$ is driven by an unusual {\em flux fractionalization} mechanism, topological in character but quite distinct from the usual well-understood Kosterlitz-Thouless mechanism~\cite{Kosterlitz_Thouless_1973,Jose_Kadanoff_Kirkpatrick_Nelson_1977} involving the proliferation of topological defects. \begin{figure}
	\centering
	\includegraphics[width=\linewidth]{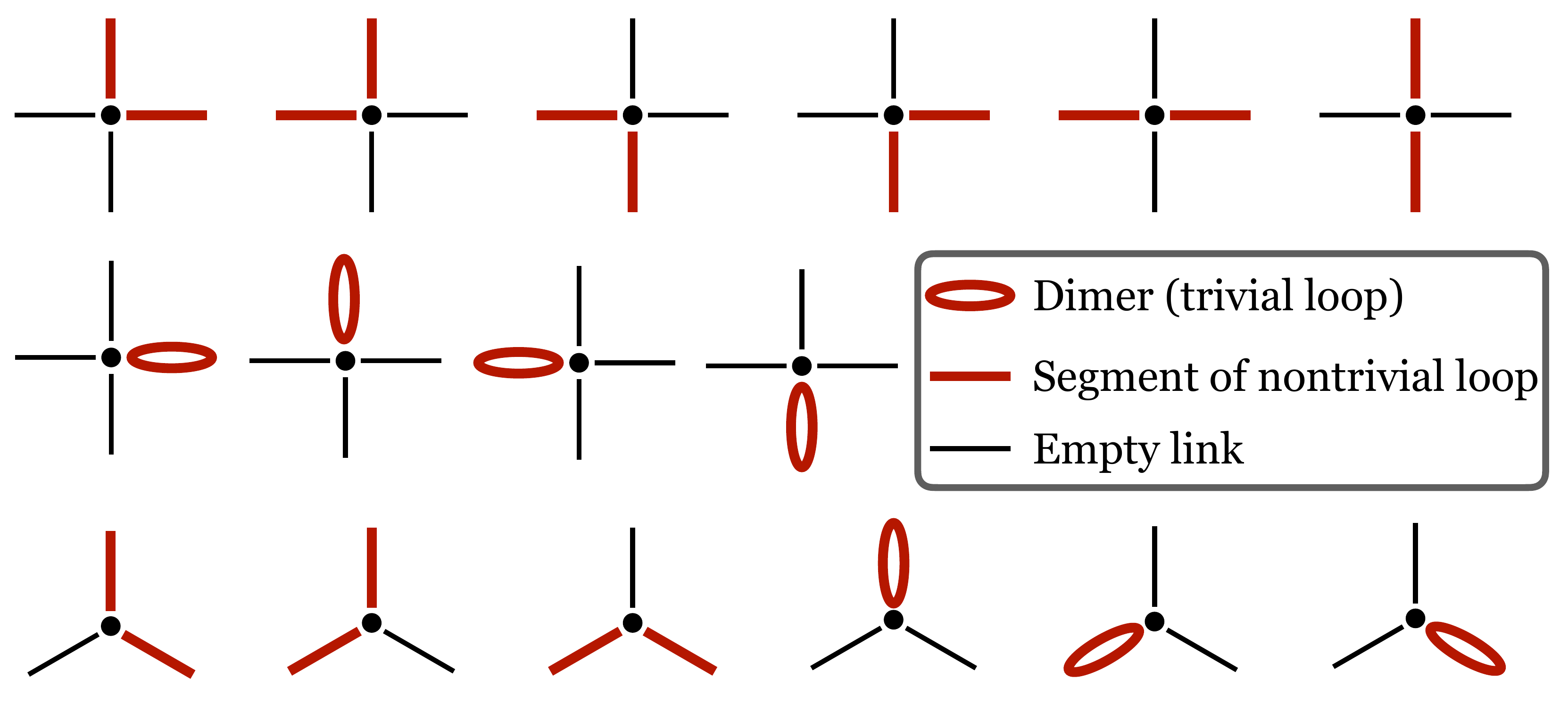}
	\caption{\label{fig:linkobjects}  Schematic representation of various ways in which a vertex of the square or honeycomb lattice can be touched by either a dimer or a nontrivial loop (of length $s \geq 4$) in a valid fully-packed configuration that contributes to the partition function $Z(w)$ defined in Eq.~\ref{eq:Zw}.}
\end{figure}


It is instructive to place this dimer-loop model and its flux-fractionalization transition in the context of the extensive statistical mechanics literature of dimer and loop models. To this end, we note that the fully-packed O(1) loop model (with each site touched by exactly one loop, and a unit fugacity for all loops) on square and honeycomb lattices is critical, with a power-law distribution of loop sizes, and power-law correlations between loop segments~\cite{Baxter_book_1989,Youngblood_Axe_1981,Blote_Nienhuis_1994,Liu_Deng_Garoni_2011}. The fully-packed dimer model on these lattices is also critical, with power-law correlations between the dimers~\cite{Kastelyn_1961,Fisher_1961,Fisher_Stephenson_1963}. These critical behaviors are both understood in terms of a long-wavelength Gaussian effective action for a scalar height field  that represents the coarse-grained electrostatic potential of fluctuating dipoles~\cite{Baxter_book_1989,Youngblood_Axe_McCoy_1980,Youngblood_Axe_1981,Henley_2010,Wilkins_Powell_2023}. These dimer and loop models are also exactly solvable~\cite{Kastelyn_1961,Fisher_1961,Fisher_Stephenson_1963}: On the square lattice, the loop model maps to the integrable six-vertex model~\cite{Baxter_book_1989}, while the partition function of the honeycomb lattice loop model is equal to that of the fully-packed dimer model since empty edges in any fully-packed dimer configuration form loops.
Fully-packed dimers on the square (honeycomb) lattice are equivalent to a system of free fermions with $\pi$ ($0$) flux on each elementary plaquette~\cite{Samuel_1980_1,Samuel_1980_2}.

These well-known  results raise the following very natural questions that do not seem to have been addressed in the classical literature on the subject:
Is there a well-motivated generalization that interpolates between these fully-packed dimer and loop models? And in such a general setting, do the dimer and loop models belong to two distinct phases, or are they just (possibly singular) limits of a single thermodynamic phase? 

Viewed from this perspective, our work answers these natural questions by constructing and studying a generalized dimer-loop model and identifying a class of anistropic frustrated magnets which could potentially provide experimental realizations of this interesting physics. 
	\begin{figure}[h]
		\includegraphics[width=\columnwidth]{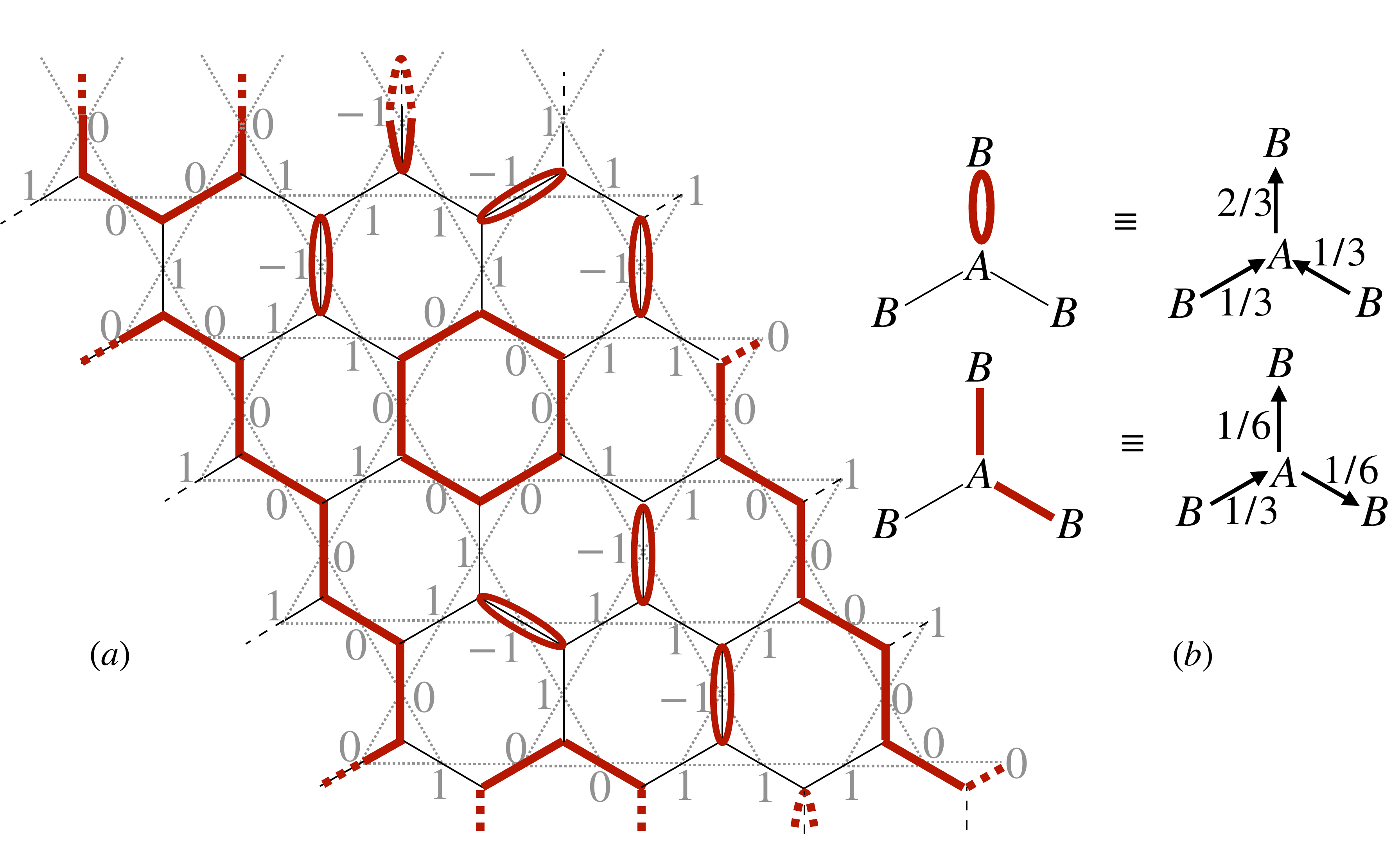}\\
		\includegraphics[width=\columnwidth]{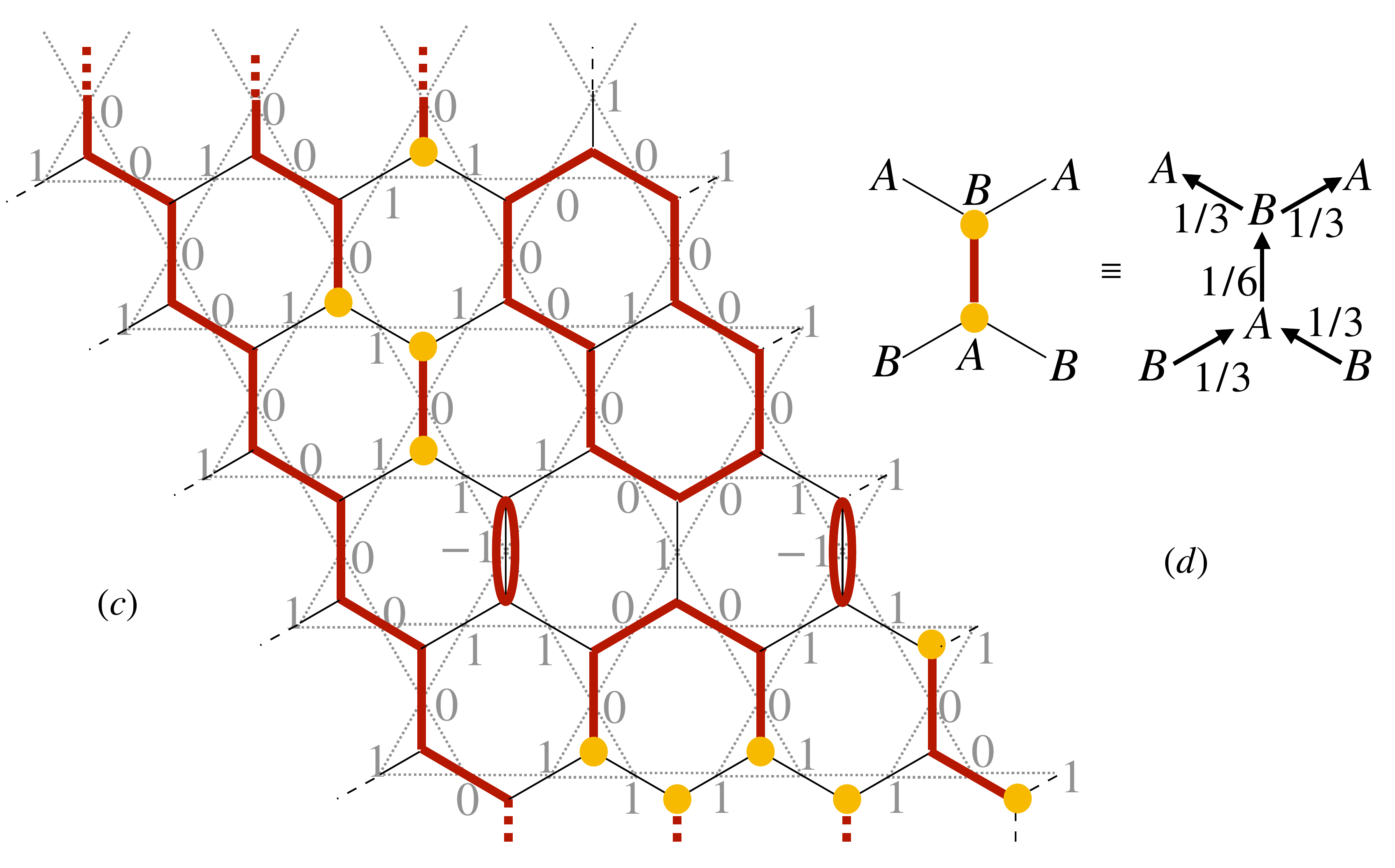}
		\caption{\label{fig:loopdimerstring}  a) An example of an allowed configuration of the dimer-loop model on a honeycomb lattice with periodic boundary conditions, drawn in a way that emphasizes that dimers are ``trivial loops'' that touch two adjacent vertices and traverse in both directions the link connecting them. Also shown is the corresponding spin configuration that contributes to the low-temperature partition function of the $S=1$ kagome magnet on its one-third magnetization plateau. b) The mapping to a divergence free polarization field is shown here. c) In the vicinity of the critical point separating the one-third magnetization plateau of the kagome magnet from its two-third magnetization plateau, the low-energy physics also has contributions from configurations with open strings as illustrated in this figure along with the corresponding spin configuration. Notice that open strings of length $s=1$ are distinct from trivial loops, since these two correspond to different local spin states. d) Each open string of any length $s \geq 1$ has a pair of charge $\pm 1/2$ defects (half-vortices) at its two ends. The fact that the polarization field develops a divergence at the locations of the half-charges is illustrated here.}
	\end{figure}

\section{Model and motivation} 
As noted in the Introduction, the classical literature on fully-packed dimer and loop models leads very naturally to questions about well-motivated generalizations that interpolate between dimer and loop models. Here, we first adopt this more theoretical perspective and introduce the dimer-loop model as an answer to these questions, and then establish its connection to the frustrated magnets of interest to us. 

As we explain in detail below, the honeycomb lattice dimer-loop model encodes the low temperature physics of a class of $S=1$ kagome magnets in an interesting magnetic field regime that corresponds to the one-third magnetization plateau of such magnets. This could potentially be realized in experimental systems. On the other hand, the square lattice dimer-loop model provides a description of the half magnetization plateau of a similar $S=1$ spin model on a planar pyrochlore lattice. Although this planar pyrochlore spin model is much less likely to be of direct relevance to an experimental system, we nevertheless provide a unified treatment of both cases, since dimer and loop models have been extensively studied on both the square and the honeycomb lattice, and their generalization studied here is therefore interesting in its own right on both lattices. Moreover, a comparison between the square and honeycomb lattice dimer-loop models is crucial for establishing the universality of various aspects of the unusual flux-fractionalization transition mechanism identified in our study.

\subsection{Definition}
\label{subsec:Definition}
In dimer models with a hard-core constraint (that forbids more than one dimer from touching any site), the dimers can also be thought of as a {\em degenerate} or {\em trivial} loops of length $s=2$, touching only two adjacent sites and traversing a single link in both directions. 
This simple observation, previously useful in other contexts~\cite{Damle_Dhar_Ramola_2012}, provides a natural theoretical motivation for the following one-parameter family of partition functions:
\begin{eqnarray}
Z(w) &=& \sum_{{\mathcal C}} w^{n_d({\mathcal C})}
\label{eq:Zw}
\end{eqnarray}
where $n_d({\mathcal C})$ is the number of such trivial loops of length $s=2$ in a fully-packed configuration ${\mathcal C}$, and the sum is over all such fully-packed configurations in which each site is touched exactly once, {\em either} by a single dimer (trivial loop of length $s=2$), {\em or} by a ``nontrivial'' loop of even length $s > 2$ (as shown in Fig.~\ref{fig:linkobjects}). In what follows, except when specifically discussing the O(1) loop model ({\em i.e.} $Z(w=0)$), we consistently use ``nontrivial loop'' to refer to loops of length $s>2$, and ``loop'' to refer to {\em any} loop,  {\em i.e.} including trivial loops of length $s=2$ (dimers). Note that this convention for assigning a loop length $s=2$ to dimers implies the following sum rule on the lengths of all loops (including dimers) in any allowed configuration: $\sum_{j=1}^{n_l} s_j = N_{\rm sites}$, where $n_l$ is the total number of loops (including dimers) and $N_{\rm sites}$ is the total number of sites of the lattice.

We reiterate that this constraint on allowed configurations is not just a theoretically natural way of connecting fully-packed hard-core dimer and loop models to each other in a more general setting. It is also forced upon us by the nature of the low-energy configurations that control the low temperature properties of certain magnetization plateaux in an interesting class of $S=1$ anisotropic magnets; this is detailed in Sec.~\ref{subsec:AnisotropicS1Antiferro}. In our work, we study $Z(w)$ on honeycomb and square lattices (corresponding respectively to magnetization plateaux of kagome and planar pyrochlore magnets) with periodic boundary conditions. It is important to note that the constraint we employ forbids not just monomers, but also disallows {\em open strings} of length $s > 1$; this is illustrated in Fig.~\ref{fig:loopdimerstring} a). Here, an {\em open string} refers to a sequence of $s$ successive occupied links that make up a simple path connecting two distinct vertices of the lattice. 


Of course, one may view the dimers of our dimer-loop model as open strings of length $s=1$ rather than trivial loops. The reader may therefore question the rationale behind our seemingly arbitrary choice of viewing dimers as trivial loops of length $s=2$ rather than open strings of length $s=1$. In fact, there are two separate reasons for our choice: First, as will be clear from the discussion in Sec.~\ref{subsec:FluctuatingDipoles}, the dimer-loop model (Eq.~\ref{eq:Zw}) on any regular bipartite lattice has a lattice-level mapping to a system of fluctuating dipoles, with each fully-packed dimer-loop configuration that contributes to $Z(w)$ mapping to a configuration of a divergence-free polarization field on the links of the lattice; this is illustrated in Fig.~\ref{fig:loopdimerstring} b). Viewing dimers as trivial loops of length $s=2$ reminds us of this mapping.
Second, as will be clear from the discussion in Sec.~\ref{subsubsec:TransitionModel}, the magnetic field driven transitions that terminate the magnetization plateaux of interest to us have a low-temperature description in terms of more general models in which open strings of all lengths $s \geq 1$ are also allowed, as illustrated in Fig.~\ref{fig:loopdimerstring} c). Note that in this more general setting (which we do not study in detail here) there are thus two objects that live on a single link of the lattice and touch the two vertices at either end of this link: a trivial loop of length $s=2$ and an open string of length $s=1$. These correspond to {\em different} local spin configurations of the $S=1$ magnet. To preserve this distinction, it is useful to view the dimers of our dimer-loop model as trivial loops.

Before we proceed with our study of the dimer-loop model defined by Eq.~\ref{eq:Zw}, some additional comments are in order: First, note that the constraint on configurations that contribute to $Z(w)$ is fundamentally different from the constraint on allowed configurations in studies of variable density dimer models that are of potential relevance to the physics of Rydberg atoms in optical lattices. For instance, in Ref.~\cite{Verresen_Lukin_Vishwanath_2021}, allowed configurations have zero or one dimer touching each vertex, while Ref.~\cite{Yan_Samajdar_2022} analyzes a model in which allowed configurations must have one or two dimers touching each vertex. The first of these allows a nonzero density of monomers that live on the vertices that are not touched by any dimer, while the second of these allows (in our language) {\em open strings of length $s>1$} in addition to nontrivial loops and trivial loops ({\em i.e.} dimers that do not touch any other dimer).  
As will be clear from the results we present, this makes the  physics of our dimer-loop models completely different from that of these variable density dimer models even if we study the latter on the same bipartite lattice as our dimer-loop model and restrict attention to the same classical regime that we focus on here.

Second, as already alluded to in the Introduction, the fully-packed $O(1)$ loop model on the honeycomb lattice is dual to the fully-packed honeycomb lattice dimer model. Indeed, there is a one-to-one correspondence that maps each configuration of the fully-packed $O(1)$ loop model to a fully-packed dimer configuration and vice versa: starting with a fully-packed loop configuration, we may place a dimer on all empty links of the original loop configuration and then delete all its loops to obtain a fully-packed dimer configuration. In the reverse direction, we obtain a unique configuration of the fully-packed $O(1)$ loop model by placing a loop segment on each empty link of a fully-packed dimer configuration, and then deleting all the dimers. Thus, on the honeycomb lattice there is an exact mapping between $Z(w=0)$ and $Z(w=\infty)$. However, we have been unable to exploit this observation to obtain a duality transformation that maps $Z(w)$ at a general $w$ to $Z(\tilde{w} = f(w))$ for some choice of $f(w)$.

Third, we emphasize that the dimer-loop model defined by $Z(w)$ also constitutes an interesting generalization of fully-packed dimer and O(1) loop models on bipartite three-dimensional lattices such as the cubic and the diamond lattice. Indeed, the diamond lattice dimer-loop model defined by $Z(w)$ provides a description of a system of $S=1$ moments on the pyrochlore lattice in an interesting regime with competing exchange and single-ion anisotropy. However, the present computational study is restricted to the two dimensional case, since the physics of the three-dimensional case is somewhat different and deserves a separate detailed study.

\subsection{Fluctuating dipoles and height mapping}
\label{subsec:FluctuatingDipoles}
To a link occupied by a segment of a nontrivial loop on any regular bipartite lattice with coordination number $z$,
we assign an ``electric polarization'' vector $\vec{E}$ of magnitude $(z-2)/2z$ pointing from the $A$ sublattice site of this link to its $B$ sublattice site. Similarly, a link occupied by a dimer is assigned an  electric polarization vector of magnitude $(z-1)/z$ pointing from the $A$ sublattice site of this link to its $B$ sublattice site. Finally, an empty link is assigned a polarization vector of magnitude $1/z$ pointing from its $B$ sublattice site to its $A$ sublattice site. Note that $z=3$ ($z=4$) for the honeycomb (square) lattice. With these assignments, the lattice divergence of the polarization vector is zero at each site in any valid fully-packed configuration. 
We write the divergence-free polarization field $\vec{E}$ as the lattice curl of a scalar field $H$ defined on sites of the triangular (square) lattice dual to the original honeycomb (square) lattice: $\vec{E} = \Delta \times H$. For a fully-packed dimer configuration without any nontrivial loops, this construction 
reduces to the usual definition of polarization fields and microscopic heights in the dimer model. 

If a nontrivial loop is cut by deleting one link to produce an open string, the configuration of $H$ develops a half-vortex of vorticity $\gamma_r /2$ at each free end $r$ of this open string, where $\gamma_r = 1$ ($\gamma_r = -1$) for $A$ ($B$) sublattice sites. 
Similarly, if a site $r$ is not touched by any dimer or nontrivial loop, it hosts a unit-strength vortex in $H$, of vorticity $\gamma_r$. Periodic boundary
 conditions on the dimers and nontrivial loops result in {\em winding} boundary conditions on $H$. In the square lattice case, the corresponding winding numbers 
$\phi_{x}$ ($\phi_{y}$) represents the flux of $\vec{E}$ across a cut spanning the system along the $x$ ($y$) axis
 of the dual lattice. In the honeycomb case, $\phi_x$ and $\phi_y$ correspond to the flux across cuts along two principal axes of the dual triangular lattice making an angle $2\pi/3$ with each other.
  The fluxes defined thus are restricted to take on half-integer values for general $w$, with an additional restriction to purely integer values that applies only at $w=\infty$.
  
This construction of a divergence-free polarization field $\vec{E}$ goes through unchanged on bipartite three-dimensional lattices such as the cubic lattice (with $z=6$) and the diamond lattic (with $z=4$). However, when we account for the zero divergence constraint, we naturally end up with a vector potential defined on links of the dual lattice, rather than a scalar height field. The physics is therefore quite different from the two-dimensional case that is our focus here, and requires a separate discussion and computational study.

\subsection{Anisotropic $S=1$ antiferromagnets}
\label{subsec:AnisotropicS1Antiferro}
We now identify and explore an interesting regime in the low-temperature physics of anisotropic $S=1$ antiferromagnets on frustrated corner-sharing lattices such as the kagome and planar pyrochlore lattices in two dimensions and the pyrochlore lattice in three dimensions. This low-energy physics is controlled by the dimer-loop models defined in the previous section and motivates their detailed study.

Our starting point is the Hamiltonian
\begin{eqnarray}
H &=& \large \sum_{\langle ij \rangle\in t } \left(J_zS^z_iS^z_j+J_{\perp}(S^x_iS^x_j+S^y_iS^y_j)\right) \nonumber \\
&& + \sum_i\Delta (S^z_i)^2-B\sum_i S^z_i
\label{eq:frustratedmagnet}
\end{eqnarray}
where $i$, $j$ are sites of a kagome (planar pyrochlore) lattice  and $ (\langle i j \rangle \in t$ refers to all links belonging to a single triangle (tetrahedron) of the kagome (planar pyrochlore) lattice, and $\vec{S}_i$ are spin $S=1$ variables. \begin{figure}
	\centering
	\subfigure[\label{}]{\includegraphics[width=0.49\linewidth]{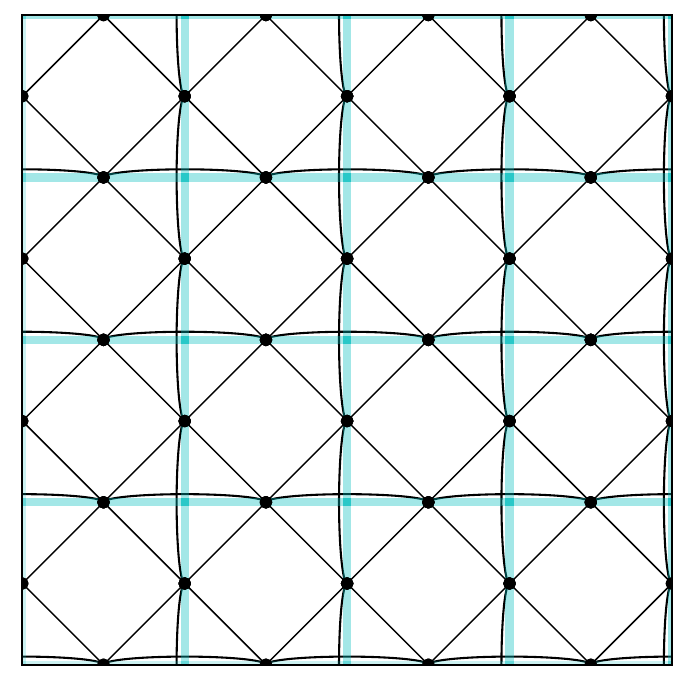}}
	\subfigure[\label{}]{\includegraphics[width=0.49\linewidth]{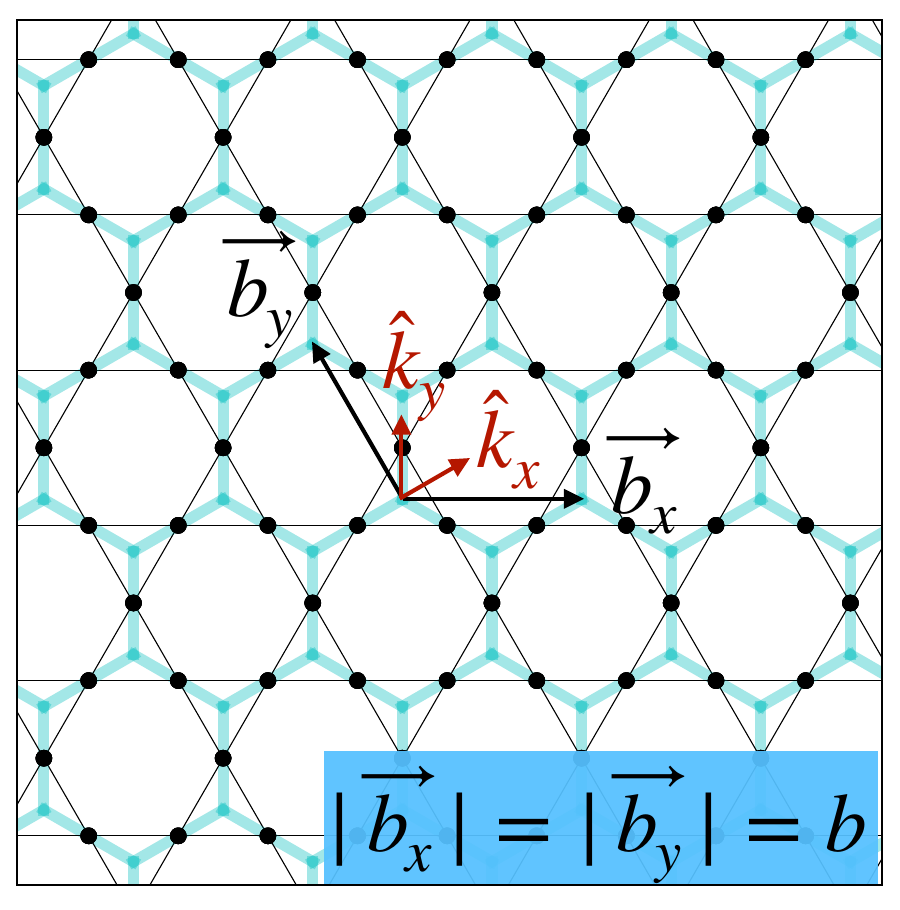}}
	\caption{\label{fig:spinlattice} (a) Planar pyrochlore lattice along with the square lattice whose links host the sites of the planar pyrochlore lattice, and whose sites are located at the centers of the tetrahedra of the planar pyrochlore lattice (b)  kagome lattice along with the honeycomb lattice whose links host the sites of the kagome lattice and whose sites are located at the centers of the triangles of the kagome lattice. In the kagome case, $\vec{b}_1$ and $\vec{b}_2$ are the translation vectors of the underlying triangular Bravais lattice, while $\hat{k}_x$ and $\hat{k}_x$ are the unit magnitude reciprocal vectors that define the conventional coordinate system in reciprocal space.}
\end{figure}

As we now show, the interplay between the dominant exchange couplings and the single ion anisotropy is particularly interesting in such corner-sharing geometries when
\begin{eqnarray}
J_z &=& J \; \; > \; \; 0 \nonumber \\
\Delta &=& J +\mu  \; ,  \nonumber \\
J_\perp & \ll & J \; ,\nonumber \\
\mu &\ll & J \; .
\end{eqnarray}
In other words, this is a regime in which there is a strong exchange anisotropy, with $J_z \equiv J$ being the dominant exchange coupling, and a {\em comparably strong} single-ion anisotropy $\Delta = J+\mu$ that favors the $m_z=0$ state of each spin $S=1$ moment. The energy scale $\mu$ then encodes the {\em relatively small} (compared to their mean value) diffference in the values of $\Delta$ and $J_z$. For temperatures $T \gg J_\perp$ and magnetic fields $B\hat{z}$ oriented in the $\hat{z}$ direction, quantum fluctuations induced by $J_{\perp}$ can be ignored to leading order since $J_\perp \ll J$, and the properties of such a frustrated antiferromagnet are controlled by the physics of an effectively classical model for $S^z_i$ variables that can take on values $\pm 1$ and $0$ at each site $i$.

\subsubsection{One-third magnetization (half magnetization) plateau on the kagome (pyrochlore and planar pyrochlore) lattice}
\label{subsubsec:PlateauModel}
For $B$ in a wide range (of ${\mathcal O}(J)$ width) around $B= (2\alpha J+\mu)$ along the $+\hat{z}$ axis, where $\alpha =1$ ($\alpha =2$) for the kagome (planar pyrochlore) lattice, the dominant ${\mathcal O}(J)$ part of the classical energy is minimized by low-energy configurations that have total spin $S^{z}_{t} = \alpha$ on each triangle (tetrahedron) $t$ of the kagome (planar pyrochlore) lattice. Indeed, each triangle (tetrahedron) that has $S^z_{t} = \alpha \pm 1$ costs ${\mathcal O}(J)$ excess energy compared to these low-energy configurations. At low temperatures $T \ll J$, this is expected to give rise to a magnetization plateau on which the total magnetization of the kagome (planar pyrocholre) magnet is frozen to  one-third (half) of its saturation magnetization.

The physics on this plateau maps directly to the clasical dimer-loop partition function for low but not-too-low temperatures $T$ in the broad range $J_\perp \ll T \ll J$. 
To see this, consider the kagome case first. In this regime on the kagome lattice, the classical configurations that contribute are those in which each triangle $t$ either has two spins taking on the value $S^z=1$ and one spin taking on the value $S^z= -1$, or one spin taking on the value $S^z=1$ and two spins taking on the value $S^z = 0$. Identifying $S^z = -1$ with a dimer on the honeycomb lattice whose links host the kagome spins (see Fig.~\ref{fig:spinlattice}), and $S^z= 0$ with a loop segment on this honeycomb lattice, we see that $Z(w)$, with $w = e^{-2\mu/T}$, maps exactly to the low temperature partition function on the magnetization plateau.  

The planar pyrochlore case is very similar. In the corresponding regime, the classical configurations that contribute to the partition function are those in which each tetrahedrom $t$ either has three spins taking on the value $S^z = +1$ and one spin taking on the value $S^z = -1$, or has two spins taking on the value $S^z = +1$ and two spins taking on the value $S^z = 0$. Identifying $S^z = -1$ with a dimer and $S^z = 0$ with a loop segment on the diamond lattice (square lattice) whose links host the spins of the pyrochlore lattice (planar pyrochlore lattice), we see that these configurations are precisely the ones that contribute to the partition function $Z(w)$ considered here. Indeed, with the identification $w= e^{-2\mu/T}$, $Z(w)$ maps exactly to the the low temperature partition function of the antiferromagnet on this magnetization plateau. 
Naturally, when the direction of the field is reversed, the kagome (planar pyrochlore) magnet has another such one-third magnetization (half-magnetization) plateau, which can also be understood in an entirely analogous way by reversing the roles of $S^z=-1$ and $S^z = +1$ in the foregoing discussion. In this paper, we focus on the positive case, with magnetization density $+\frac{1}{3}$ ($+\frac{1}{2}$) per site along the $\hat{z}$ axis in the kagome (planar pyrochlore) case.

As the temperature is varied in the range $J_{\perp} \ll T \ll J$, the coupling constant $w$ sweeps through a range of values that is determined by the sign of $\mu$: For either sign of $\mu$, $w \approx 1$ for high temperatures in the range $J \gg T \gg |\mu|$. When $\mu$ is negative, $w$ increases rapidly from this ${\mathcal O}(1)$ value as the temperature is lowered, while for positive $\mu$, $w$ decreases rapidly to zero as the temperature is lowered. Thus, the low temperature physics on the magnetization plateau of a given frustrated system will be controlled by properties of the dimer-loop model over the corresponding range of $w$.

\subsubsection{Transition to the two-third (three-fourth)  magnetization plateau on the kagome (pyrochlore and planar pyrochlore) lattice}
\label{subsubsec:TransitionModel}
Interestingly, the transition that terminates this one-third magnetization (half magnetization) plateau at its high-field end on the kagome (planar pyrochlore) lattice is described by an extended dimer-loop model in which unit-vortices continue to be forbidden, but there is a nonzero fugacity for half-vortices.  To see this, we first note the following: As the field is increased further, this plateau terminates when it becomes more energetically favorable for each triangle (tetrahedron) of the kagome (planar pyrochlore) lattice to have a total spin $S^z_{t} = \alpha +1$, with $\alpha$ defined as before. The corresponding energetic threshold is found to be at $B_{\rm upper} = (2\alpha + 1) J+\mu$.  

Any triangle (tetrahedron) $t$ of the kagome (planar pyrochlore) lattice with  $S^z_t = \alpha +1$ must have exactly one spin with $S^z = 0$ and all other spins with $S^z=+1$. This falls outside the configuration space of the dimer-loop model in which each site is touched by exactly one loop of even size $s \geq 2$. Indeed, it corresponds to a half-vortex in the height field, with vorticity $\pm 1/2$ (associated with a divergence of the polarization field at its location). This half-vortex terminates an open string of length $s \geq 1$ which has another half-vortex at its other end. Thus, this enlarged space of configurations now allows open strings of length $s\geq 1$ (with half-vortices at their ends) in addition to trivial loops (of size $s=2$) and nontrivial loops (of even size $s>2$). This is illustrated in Fig.~\ref{fig:loopdimerstring} c) and d). However, unit-vortices remain forbidden at low temperature in this regime due to an ${\mathcal O}(J)$ energy cost. The nature of this field-driven transition to the higher magnetization plateau is therefore controlled by the physics of this more general dimer-loop model in which unit vortices are forbidden but there is a nonzero fugacity of half-vortices.

In the vicinity of this transition, we parameterize $B = B_{\rm upper} +  \epsilon$, with $\epsilon \ll J$. With this parameterization, the low-temperature physics in this vicinity is modeled by the partition function
\begin{eqnarray}
Z(w, f_{\frac{1}{2}}) &=& \sum_{{\mathcal C}} w^{n_d({\mathcal C})} f_{\frac{1}{2}}^{n_h({\mathcal C})} \; ,
\label{eq:Zwf1/2}
\end{eqnarray}
where the sum is now over all dimer-loop configurations ${\mathcal C}$ in which unit-vortices are forbidden but half-vortices (with vorticity $\pm 1/2$) are allowed, $n_{h}$ is the number of half-vortices, 
$w = \exp(-2 \mu/T)$ as before, 
and $f_{\frac{1}{2}}= \exp(\epsilon/2T)$. 
The physics of this extended dimer-loop model with $f_{\frac{1}{2}}\neq 0$, which models the transition between plateaus, is also expected to be quite interesting, and deserves a separate computational study. This is discussed further in Sec..~\ref{sec:Outlook}

Finally, we re-emphasize a point that was already mentioned briefly in Sec.~\ref{subsec:Definition}: This extended dimer-loop model has a configuration space that allows two distinct objects that occupy a single link of the lattice and touch the two vertices that it connects: open strings of length $s=1$, and trivial loops (that are assigned a length $s=2$ in our convention). A trivial loop corresponds to an {\em isolated} spin $S^z = -1$ (with all neighboring spins on the medial lattice taking on the value $S^z=+1$), while an open string of length $s=1$ represents an {\em isolated} $S^z = 0$ on the corresponding site of the medial lattice. This is very different from the configuration space of the variable density dimer model studied in Ref.~\cite{Yan_Samajdar_2022}, which has only one kind of object that occupies a single link of the lattice. In addition, the classical limit of this variable density dimer model has Boltzmann weights that are very different from the weights that appear in Eq.~\ref{eq:Zwf1/2}. The low-temperature behavior encoded in $Z(w, f_{\frac{1}{2}})$ is therefore expected to be completely different.

\subsubsection{Possible extension to three dimensions}
The foregoing identificaton of the low-energy models that control the behavior on the half-magnetization plateau and its transition to the three-fourth magnetization plateau in the planar pyrochlore case also applies essentially without change to the three-dimensional pyrochlore case; the corresponding dimer-loop model and its extension lives on the bipartite diamond lattice. The fully-packed hard-core dimer model and the fully-packed $O(1)$ loop model on such bipartite lattices in three dimensions both have interesting long-distance physics~\cite{Huse_Krauth_Moessner_Sondhi_2003,Nahum_Chalker_Serna_etal_2013_PRL}, although the effective field theory in three dimensions is rather different from the two-dimensional case~\cite{Henley_Coulombphasesreview2010}. This raises interesting questions about the long-distance physics encoded in the diamond lattice version of $Z(w)$. 

In the remainder of this work, we focus on the physics of the one-third (one-half) magnetization plateau of the kagome (planar pyrochlore) magnet, leaving these interesting questions about the three-dimensional case to future work.

\section{ Monte-Carlo algorithm and measurements} 
\label{sec:AlgoAndMeasurements}
Here, we study the properties of this dimer-loop model using Monte Carlo (MC) simulations as a function of $w$ for periodic square and honeycomb lattices with $L \times L$ unit cells with $L$ ranging from $128$ to $2048$ ($96$ to $1536$) for the square (honeycomb) lattice ($L$ is chosen to be a multiple of $12$ for the honeycomb lattice case and a power of $2$ for the square lattice). Our MC simulations employ two different variants of a worm update~\cite{Sandvik_Moessner_2006,Alet_Ikhlef_Jacobsen_etal_2006}, the ``half-vortex update'' and the ``unit-vortex update''. The half-vortex update is designed using ideas from Ref.~\cite{Rakala_Damle_2017} to maintain detailed balance in a larger configuration space which has a half-vortex in the microscopic height field $H$ at the (fixed) tail of the worm, and another half-vortex of opposite charge attached to the head of the worm that moves with detailed balance. The unit-vortex update employs a worm construction that creates a unit-vortex at the fixed tail of the worm and a unit vortex of opposite charge attached to the head of the worm that moves with detailed balance; by construction, these unit vortices are constrained to maintain their integrity and not break up into half vortices.

\subsection{Half-vortex worm update}
\label{subsec:halfvortexwormupdate}
The half-vortex worm update starts at a random site $v_0$. This is the first {\em entry site} $e_0$, at which the ``worm tail'' is held fixed during worm construction (motivation for the nomenclature will become clear from the rest of this description). If $v_0$ is touched by a dimer connecting it to a neighbor $v_1$, we either abort the construction with probability $1/2$, or go to $v_1$ with probability $1/2$. In the latter case, $v_1$ becomes the first {\em pivot site} $\pi_0$. If $v_0$ is touched by a nontrivial loop, we randomly choose (with probability $1/2$ each) to go to one of the two neighbors of $v_0$ connected to it by segments of this loop. This chosen neighbor becomes the first pivot site $\pi_0$. \begin{figure}
	\centering
	\includegraphics[width=\linewidth]{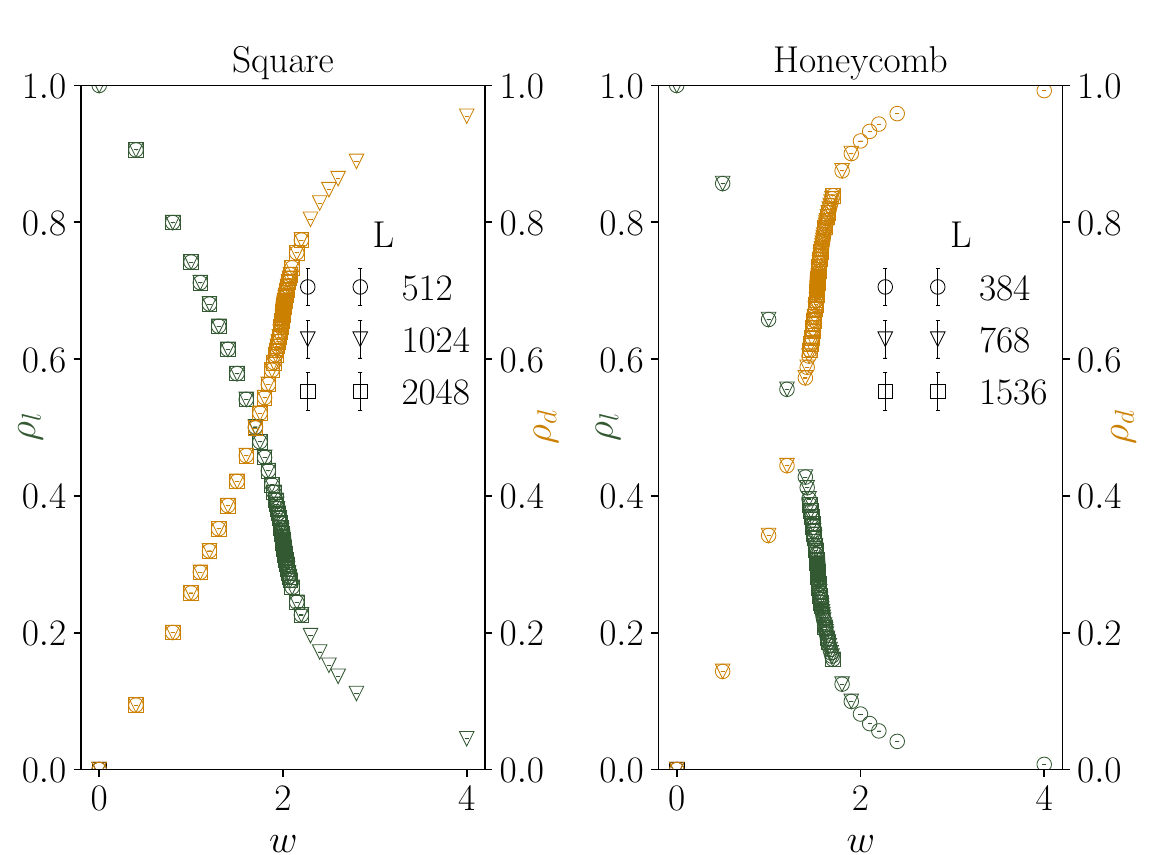}
	\caption{\label{fig:rhosandlvsw} The densities $\rho_d$ and $\rho_l$, of sites touched by a dimer or a loop respectively, are both nonzero for any finite $w>0$ and vary continuously with $w$.}
\end{figure}

The rest of the worm update proceeds as follows. At each pivot $\pi_n$ reached from entry $e_n$, we pivot (one of) the loop segment(s) originally connecting $\pi_n$ to $e_n$, so that this loop segment now connects $\pi_n$ to the exit site $e'_{n}$, chosen from among its neighbors using a probability table that satisfies detailed balance. Thus, at this step, if $e_n$ and $\pi_n$ are connected by a dimer, this dimer is converted into a segment of a nontrivial loop, which has another segment connecting $\pi_n$ to the exit $e'_n$. If on the other hand $e_n$ and $\pi_n$ are connected by a segment of a nontrivial loop, that segment is pivoted around $\pi_n$ so that it now connects $\pi_n$ to $e'_n$, and $e_n$ is no longer connected to $\pi_n$ by a loop segment. In this way, the ``worm head'' now reaches $e'_{n}$. 

In general, at intermediate steps, the exit $e'_n$ reached in this way either has another nontrivial loop already touching it, or a dimer already touching it. In the former case, we go along this nontrivial loop in one of the two possible ways (with probability $1/2$ each) to reach one of the two neighbors connected to $e'_n$  by this loop. This neighbor becomes the new pivot $\pi_{n+1}$, which has been entered via the new entry site $e_{n+1} \equiv e'_n$. If $e'_n$ was originally touched by a dimer, we go to the other end of this dimer, which becomes the new pivot $\pi_{n+1}$, reached from the new entry site $e_{n+1} \equiv e'_n$. 
This continues until the current exit $e'_n$ is not already touched by another nontrivial loop or by a dimer. When this happens, $e'_n$ necessarily coincides with $v_0$ (the randomly chosen starting site), and the half-vortex that was moving with the worm head gets annihilated by the half-antivortex that was held fixed at the worm tail. This results in a new allowed configuration which can be accepted with probability $1$. \begin{figure}
	\centering
	\includegraphics[width=\linewidth]{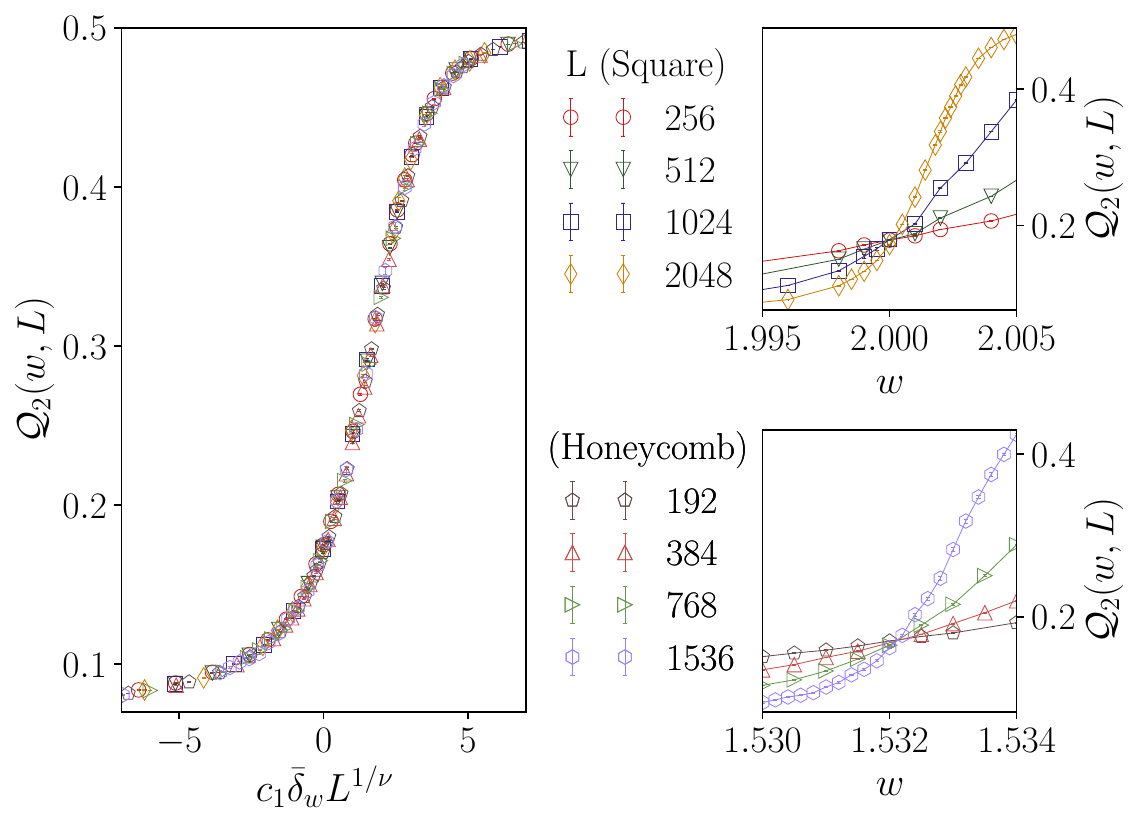}
	\caption{\label{fig:Qvsw} Right panels: The Binder ratio ${\mathcal Q}_2$ of loop sizes, defined in Sec.~\ref{sec:AlgoAndMeasurements}, shows a clear crossing at a critical value $w_c \approx 2.000(1)$ ($w_c \approx 1.5321(2)$) on the square (honeycomb) lattice. Left panel: Data for $w$ close to $w_c$ for various sizes $L$ collapses on to the scaling form described in Eq.~\ref{eq:Q2AndPScaling}. The scaling collapse displayed here employs the following parameter values: $w_c = 2.000$ ($w_c = 1.5321$) for the square (honeycomb) lattice case, $\nu=1.0$ for both cases, and  $c_1=1$ ($c_1 = 1.68$) for the square (honeycomb) lattice. Note that the choice $c_1 =1$ for the square lattice case is a convention that defines the scaling function $F_{{\mathcal Q}}$ from the collapsed square lattice data. }
\end{figure}

\subsection{Unit-vortex worm update}
\label{subsec:unitvortexwormupate}
The unit-vortex worm update is simpler to describe: One simply deletes all sites that are touched by nontrivial loops of a configuration (and all links that connect these deleted sites to the rest of the lattice). This depleted lattice hosts a fully-packed dimer cover. We now use the standard dimer worm algorithm~\cite{Sandvik_Moessner_2006,Alet_Ikhlef_Jacobsen_etal_2006} on this depleted lattice to obtain a new fully-packed dimer cover, which can be acccepted with probability $1$. This update is not useful or efficient for $w \ll w_c$, but plays a crucial role in equilibrating the system for $w>w_c$. 

\subsection{Measurements}
\label{subsec:measurements}
As a result of the manner in which these worm updates are constructed, the measured histograms of the head-to-tail displacements in these two worm updates are proportional to the corresponding correlation functions $C_{v}^{q}(\vec{r})$ ($q=1/2,1$) for a pair of test vortices with vorticity $\pm q$ ($q=1/2,1$). In the unit vortex case, the constraint that each unit vortex maintains its integrity and does not break up into half vortices is important, and the measured histograms therefore need to be interpreted appropriately (this is discussed further in Sec.~\ref{subsec:offcriticaltheory}). \begin{figure} 
	\centering
	\includegraphics[width=\linewidth]{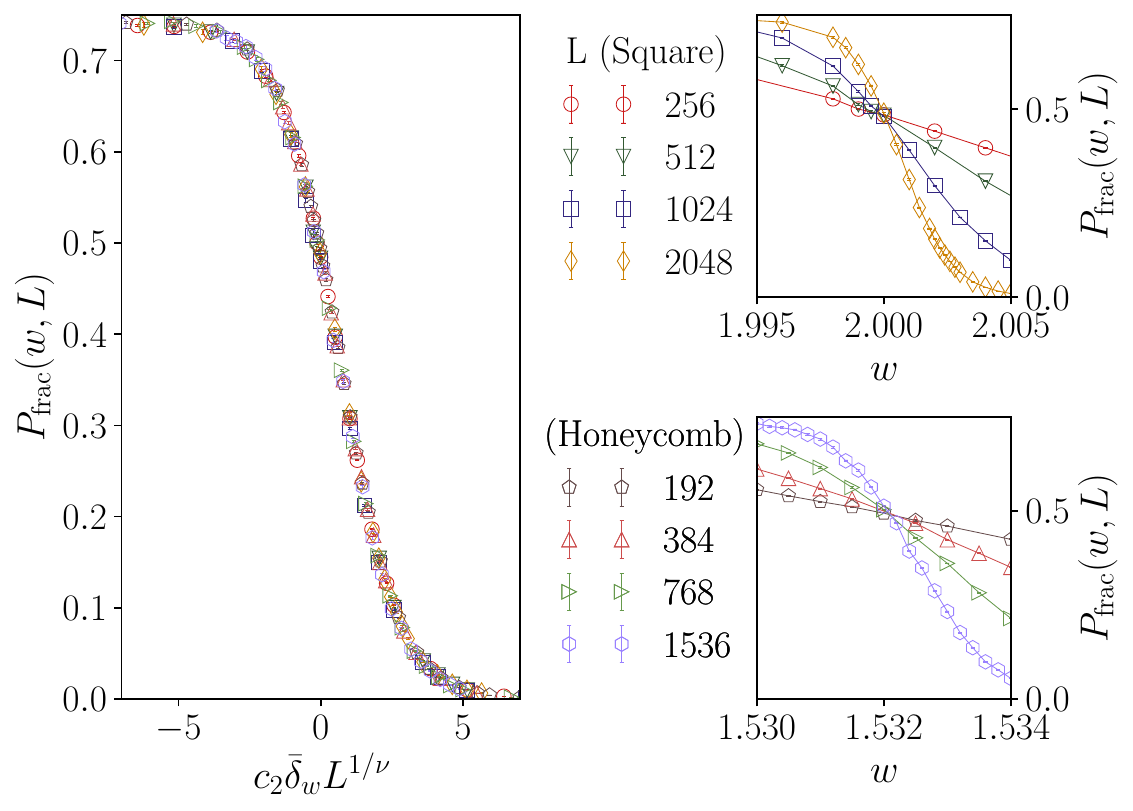}
	\caption{\label{fig:Pintvsw}  Right panels: The probability of finding fractional fluxes $P_{\rm frac}$, defined in Sec.~\ref{subsec:Loops&Fluxes}, shows a clear crossing at a critical value $w_c \approx 2.000(1)$ ($w_c \approx 1.5321(3)$) on the square (honeycomb) lattice. Left panel: Data for $w$ close to $w_c$ for various sizes $L$ collapses on to the scaling form described in Eq.~\ref{eq:Q2AndPScaling}. The scaling collapse displayed here employs the following parameter values: $w_c = 2.000$ ($w_c = 1.5321$) for the square (honeycomb) lattice case, $\nu=0.999$ for both cases, and  $c_2=1$ ($c_2 = 1.69$) for the square (honeycomb) lattice. Note that the choice $c_2 =1$ for the square lattice case is a convention that defines the scaling function $F_{P}$ from the collapsed square lattice data.}
\end{figure}

We also measure the density $\rho_d$ and $\rho_l$ of sites touched respectively by dimers and nontrivial loops (with the normalization $\rho_d+\rho_l = 1$ independent of $w$), the loop size distribution $P_l(s,L)$ and associated moments $S_m = \langle \sum_{j=1}^{n_l} s_j^m \rangle$ ($m=2,4$), as well as the ratios $R = S_4/S_2^2$, ${\mathcal Q}_2 = \langle \sum_{i \neq j} s_i^2 s_j^2\rangle/S_2^2$, and the joint flux distribution $P(\phi_x,\phi_y)$ and its marginals $P_{x} = \sum_{\phi_{y}} P(\phi_x,\phi_y)$ (and similarly for $P_y$) as a function of $w$ and $L$. We also measure the correlation function $C_{\psi}$ of the local columnar order parameter field. On the square lattice, this is defined using the conventions of Ref.~\cite{Ramola_Damle_Dhar_2015} by exploiting the identification between nontrivial loops of length $s=4$ in the present model and hard squares in the lattice gas of Ref.~\cite{Ramola_Damle_Dhar_2015}. On the honeycomb lattice, this is the local field that corresponds to ordering at the three-sublattice wavevector of underlying triangular Bravais lattice. In addition, since spin-polarized neutron scattering is expected to be sensitive to the $zz$ component of the spin structure factor of the kagome magnet, we also use our data to compute this structure factor.

\section{Results}
\label{sec:Results}

\subsection{Fluxes and loop sizes}
\label{subsec:Loops&Fluxes}

From Fig.~\ref{fig:rhosandlvsw}, we see that $\rho_d$ and $\rho_l \equiv 1-\rho_d$ are both nonzero for all finite nonzero $w$,  and vary continuously with $w$. Nevertheless, there are two distinct phases, separated from each other by a critical point at a nonzero value of $w$.  \begin{figure}
	\centering
	\includegraphics[width=\linewidth]{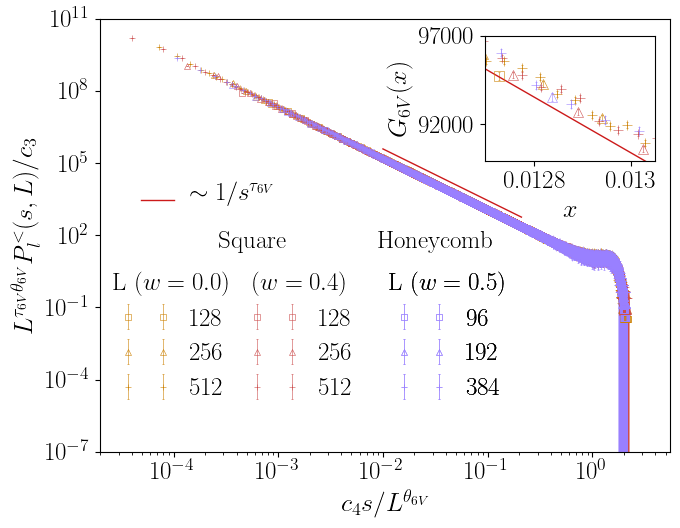}
	\caption{\label{fig:Plcollapse} Data for the loop size distribution $P^{<}_{l}(s,L)$ at various sizes $L$ for all
	 $w<w_c$ obeys the scaling form defined in Eq.~\ref{eq:P<}, with exponents $\theta_{6V}$ and $\tau_{6V}$ obtained from the exact solution of the six-vertex model on the square lattice. With the convention that $c_3=c_4=1$ for the collapse of the $w=0$ data on the square lattice, this is illustrated in the figure using data for $w=0.4$ on the square lattice and $w=0.5$ on the honeycomb lattice. These data sets are seen to obey the scaling form to high accuracy with the
	  choices $c_4=1.10$, $1/c_3=1.88$ in the former case, and $c_4=0.594$, $1/c_3 = 3.917$ in the latter case.}
\end{figure}

This is immediately apparent from the $w$ dependence of $R$ and ${\mathcal Q}_2$, and from the $w$ dependence of $P_{\rm frac} = 1-\sum_{\phi_x \in Z, \phi_y \in Z} P(\phi_x,\phi_y)$, the
 probability that  at least one out of $\phi_x$ and $\phi_y$ takes on a fractional value. In Fig.~\ref{fig:Qvsw} and Fig.~\ref{fig:Pintvsw},
  we see that curves corresponding to ${\mathcal Q}_2$ and $P_{\rm frac}$ for different $L$ cross at a well-defined critical threshold $w_c =2.000(1)$ ($w_c =1.5321(3)$) for the square (honeycomb) lattice.  Indeed, ${\mathcal Q}_2(w,L)$ and $P_{\rm frac}$ both collapse onto universal critical scaling forms
\begin{eqnarray}
{\mathcal Q}_2 (w,L) &=& F_{{\mathcal Q}}(c_1\bar{\delta}_w L^{1/\nu}) \nonumber \\
P_{\rm frac}(w,L) &=& F_{P}(c_2\bar{\delta}_w L^{1/\nu})
\label{eq:Q2AndPScaling}
\end{eqnarray}
with $\bar{\delta}_w = (w-w_c)/w_c$, $\nu \approx 1.00(1)$ and lattice-dependent scale factors $c_1$ and $c_2$. $R$ is also found to scale in this manner with the same values of $\nu$ and $w_c$ within errors, but a different scaling function and scale factors.

The $w$ dependence of $P_{\rm frac}$ demonstrates that this transition at $w_c$ is a {\em flux fractionalization transition} from the perspective of the dimer model at $w=\infty$: fractional fluxes survive in the thermodynamic limit for $w< w_c$, but not for $w> w_c$. Interestingly, the restriction to integer flux sectors for $w>w_c$ is an emergent one, in the sense that it an does {\em not} directly follow from nature of the allowed configurations at a microscopic level. Indeed, at a microscopic level, such a restriction only exists at $w=\infty$, {\em i.e.} when the only contribution to $Z(w)$ comes from fully-packed dimer configurations (without any nontrivial loops). Nevertheless, as is clear from Fig.~\ref{fig:Pintvsw}, only integer flux sectors contribute to $Z(w)$ in the thermodynamic limit for {\em all} $w>w_c$.  \begin{figure}
	\centering
	\includegraphics[width=\linewidth]{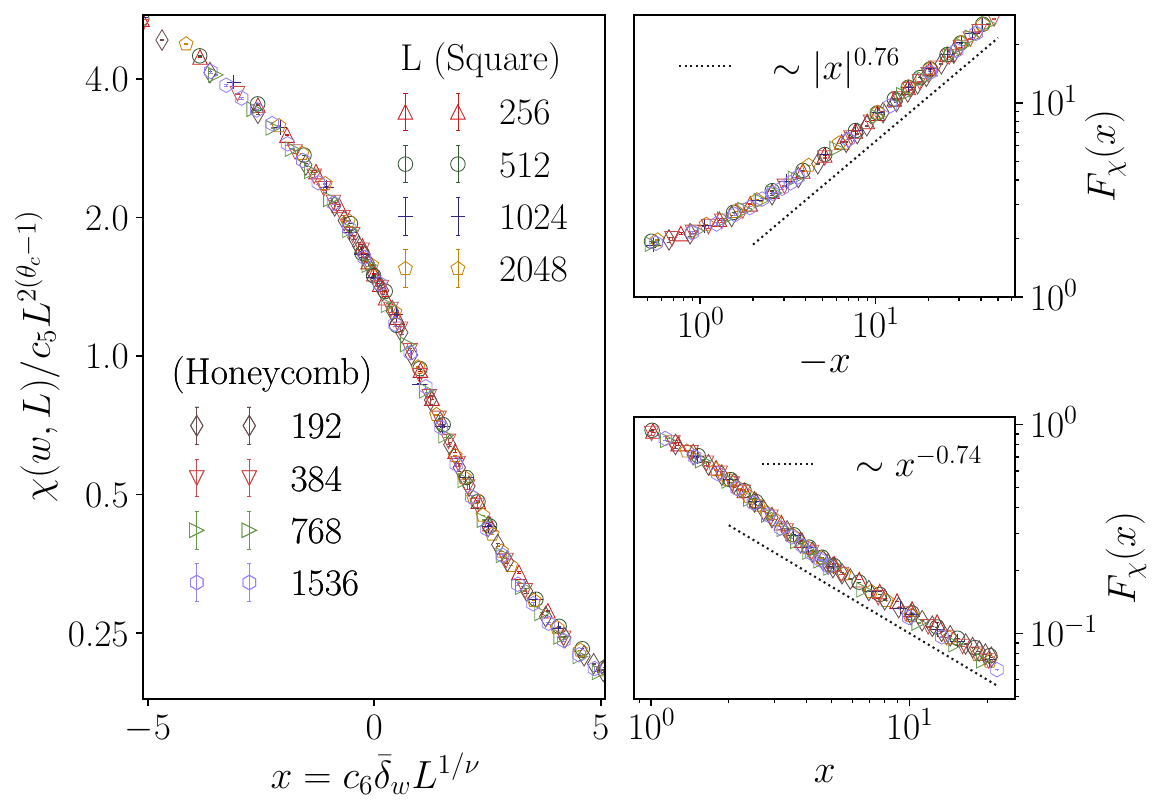}
	\caption{\label{fig:chiscaling} The loop size susceptibility on both the square and the honeycomb lattice is seen to obey the scaling form of Eq.~\ref{eq:chi}, with a common choice of $\theta_c=1.37$ and $\nu=1.0$ for the exponents, and $w_c=2.000$ ($w_c=1.5321$) for the square (honeycomb) lattice. The displayed data collapse corresponds $c_5=c_6=1$ (by convention) for the square lattice, and $1/c_5 = 0.337$, $c_6=1.69$ for the honeycomb lattice.}
\end{figure}


For all $w < w_c$ on both lattices, we also see from Fig.~\ref{fig:Plcollapse} that $P_l(s,L)$ collapses on to a universal scaling form 
\begin{eqnarray}
P_l^{<}(s,L) = \frac{c_3}{L^{\tau_{6V} \theta_{6V}}} G_{6V}\left(\frac{c_4 s}{L^{\theta_{6V}}}\right)
\label{eq:P<}
\end{eqnarray}
with  lattice and $w$ dependent constants $c_{3/4}$, and universal exponent values~\cite{Saleur_Duplantier_1987,Kondev_Henley_1995_PRL,Jacobsen_Kondev_1998,Jaubert_Haque_Moessner_2011} $\tau_{6V} = 15/7 \equiv 2.142857\dots$ and $\theta_{6V} = 7/4 \equiv 1.75$ characteristic of the exactly solvable six vertex model  that maps to the $w=0$ point on the square lattice. For $x \ll 1$, $G_{6V}(x) \sim x^{-\tau_{6V}}$. And for $x \gg 1$, $G_{6V}(x)$ vanishes rapidly with increasing $x$. The latter reflects the fact that the largest loop length scales as $s_{\rm max} \sim L^{\theta_{6V}}$, while the former corresponds to a power-law distribution of loop sizes $P^{<}_l(s,L)\sim 1/s^{\tau_{6V}}$ for $1 \ll s \ll L^{\theta_{6V}}$. \begin{figure} 
	\centering
	\includegraphics[width=\linewidth]{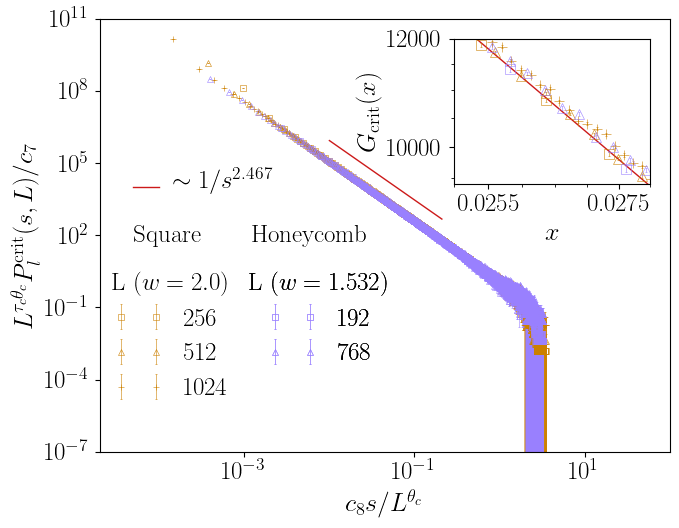}
	\caption{\label{fig:critloopsize} Data for the critical loop size distribution $P^{\rm crit}_{l}(s,L)$ at $w_c=2.000$ ($w_c=1.5320$) for the square (honeycomb) lattice for various sizes $L$ on both lattices obeys the scaling form defined in Eq.~\ref{eq:Pcrit}, with common exponent choices $\theta_c = 1.375$ and $\tau_c =2.467$ for both lattices. Here, with the convention that $c_7=c_8=1$ for the collapse of the square lattice data set, the displayed collapse on the honeycomb lattice corresponds to $1/c_7 = 1.57$ and $c_8=0.614$.}
\end{figure}

Before we proceed, it is useful to note that $\tau_{6V} \theta_{6V} = \theta_{6V} +2$. This is a consequence of a scaling relation between the exponents $\theta$ and $\tau$, that can be understood as follows: A scaling form of the type used in Eq.~\ref{eq:P<} encodes both the power-law scaling of the loop size distribution in the thermodynamic limit as well as the fact that the largest loop in a sample of linear size $L$ scales as $L^{\theta}$. Let us now suppose that the {\em number} of loops with length $s \gtrsim L^{\theta}$ scales as $L^{\zeta}$ for a sample of linear dimension $L$. If this is so, one must have
\begin{eqnarray}
L^{\zeta} &\sim & L^2 \int_{L^{\theta}}^{L^2} P_l^{<}(s,L)ds \; ,
\label{eq:scalinglaw1}
\end{eqnarray}
where the prefactor of $L^2$ multiplying the integral on the right hand side accounts for the fact that the total number of loops $n_l$ scales as $n_l \sim L^2$ for a sample of linear size $L$.
Using the scaling form Eq.~\ref{eq:P<} for $P_l^{<}(s,L)$ and taking the limit of large $L$, we see that this implies that
\begin{eqnarray}
\tau \theta  &=& 2+\theta -\zeta
\label{eq:scalinglaw2}
\end{eqnarray}
If $\zeta = 0$, {\em i.e.} if there are only $O(1)$ loops of size $s \gtrsim L^{\theta}$ in a sample of linear size $L$, this reduces to $\tau \theta = 2+\theta$. And indeed, we can verify that the ratio of the second largest loop length to the largest loop length scales to zero in the thermodynamic limit for $w<w_c$, implying that $\zeta_{6V} = 0$, consistent with the fact that $\tau_{6V} \theta_{6V} = \theta_{6V} +2$. This is a simpler, and perhaps more transparent (since it provides an operational definiton of $\zeta$ in terms of properties of the scaling function for the loop size distribution),  variant of the scaling argument given in  Ref.~\cite{Kondev_Henley_1995_PRL} for contour lines of a Gaussian free field. \begin{figure}
	\centering
	\includegraphics[width=\linewidth]{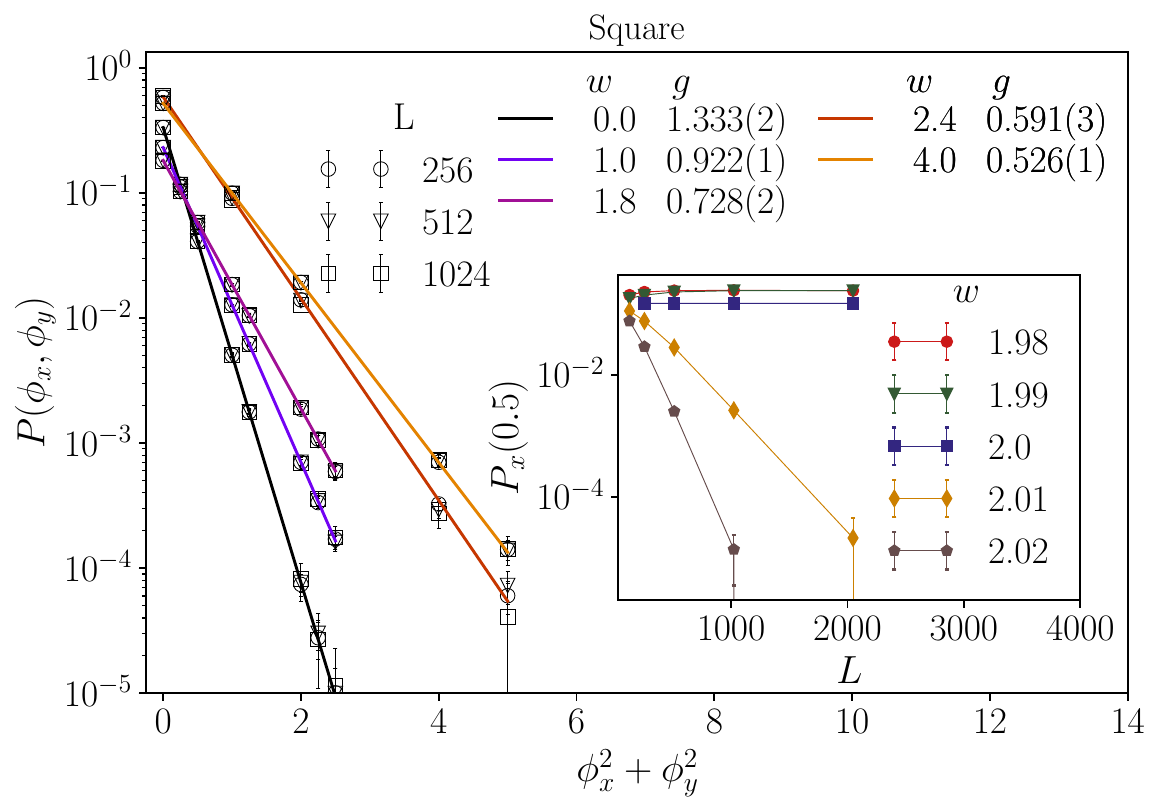}
    \includegraphics[width=\linewidth]{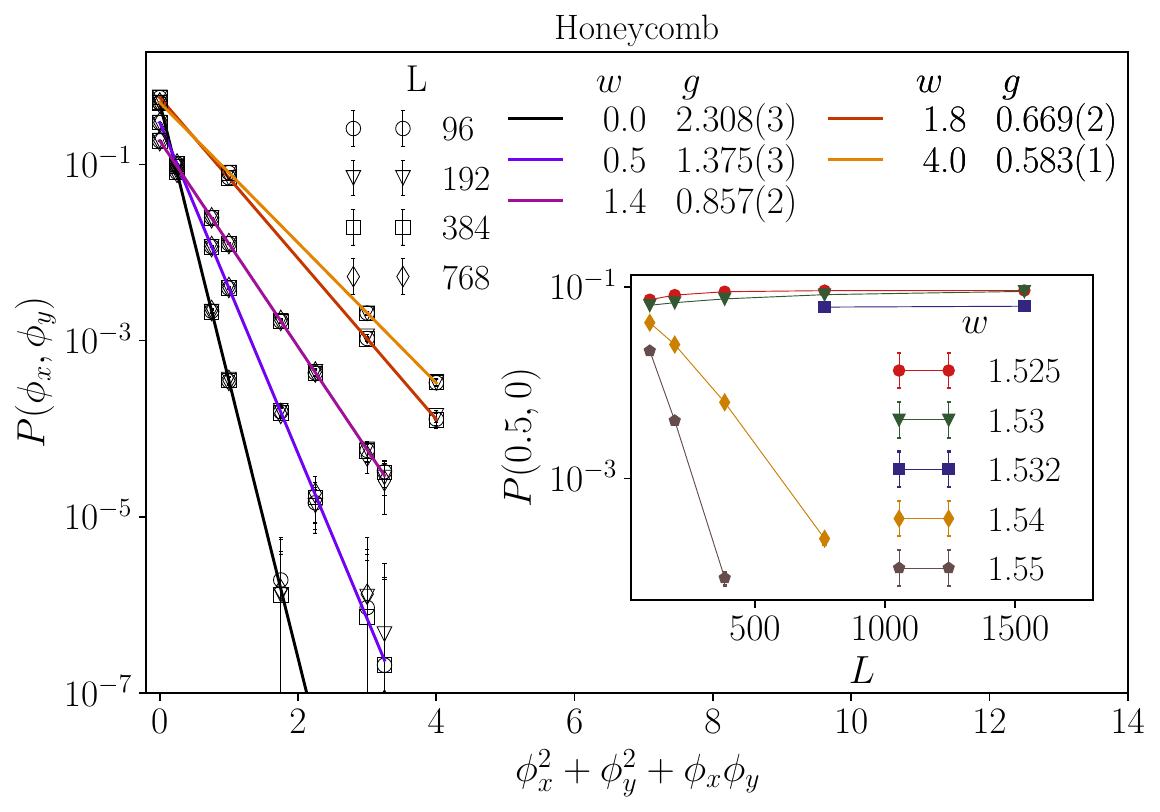}
	\caption{\label{fig:fluxsquareandhoneycomb} In both phases, the joint distribution of integer-valued fluxes $\phi_x$ and $\phi_y$ is seen to be of the $L$-independent Gaussian form given by Eq.~\ref{eq:Pjointsquare} in the square lattice case and Eq.~\ref{eq:Pjointhoneycomb} in the honeycomb lattice case. This is illustrated by the top panel for the square lattice and the bottom panel for the honeycomb lattice. For $w<w_c$, fractional values of the fluxes are governed by the same $L$-independent Gaussian distribution as integer values of the fluxes. In contrast, fractional values of the fluxes have vanishingly small probability in the thermodynamic limit for $w>w_c$, and are not governed by this Gaussian distribution. This is illustrated for both lattices by the data shown in the insets of the respective panels.}
\end{figure}

In this small-$w$ phase, the loop susceptibility $\chi = S_2/L^2$ scales as $\chi \sim L^{2(\theta_{6V}-1)}$, being dominated by the largest loops. For $w > w_c$, we find $\chi \sim O(1)$ at large $L$, as befits a short-loop phase dominated by dimers. Interestingly, at $w=w_c$, we find $\chi \sim L^{2(\theta_c -1)}$ with $\theta_c \approx 1.375(10)$, {\em i.e.} rather different from $\theta_{6V} \equiv 1.75$. Indeed, in Fig.~\ref{fig:chiscaling}, we see that all our data for  $\chi(w,L)$ for both lattices in the vicinity of the respective critical points collapses onto a universal scaling form
\begin{eqnarray}
\chi(w,L) &=& c_5L^{2(\theta_c -1)} F_{\chi}(c_6 \bar{\delta}_w L^{1/\nu})
\label{eq:chi}
\end{eqnarray}
with $\nu \approx 1.00(1)$. For large negative $x$, $F_{\chi}(x) \sim |x|^{p}$, where $p \equiv 2\nu(\theta_{6V}-\theta_c) \approx 0.75(2)$, while for $x \gg 1$, $F_{\chi}(x) \sim x^{-p'}$, where $p' \equiv 2\nu(\theta_c-1) \approx 0.75(2)$.  \begin{figure}
	\centering
	\subfigure[\label{}]{\includegraphics[width=0.49\linewidth]{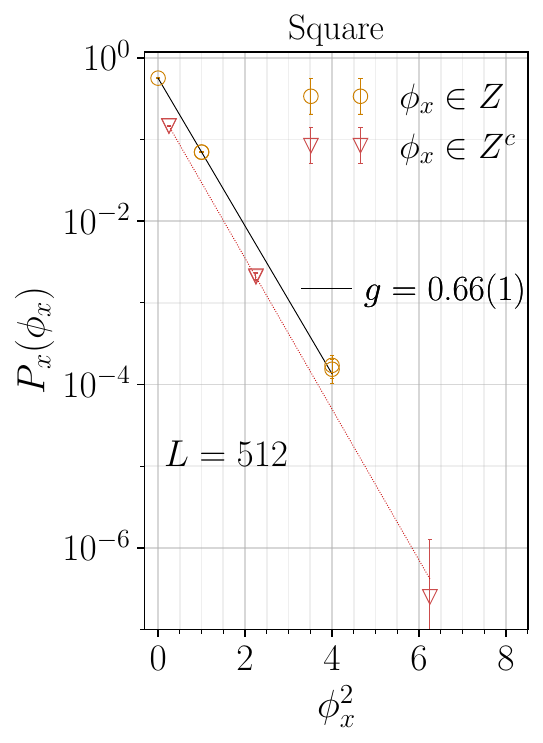}}
	\subfigure[\label{}]{\includegraphics[width=0.49\linewidth]{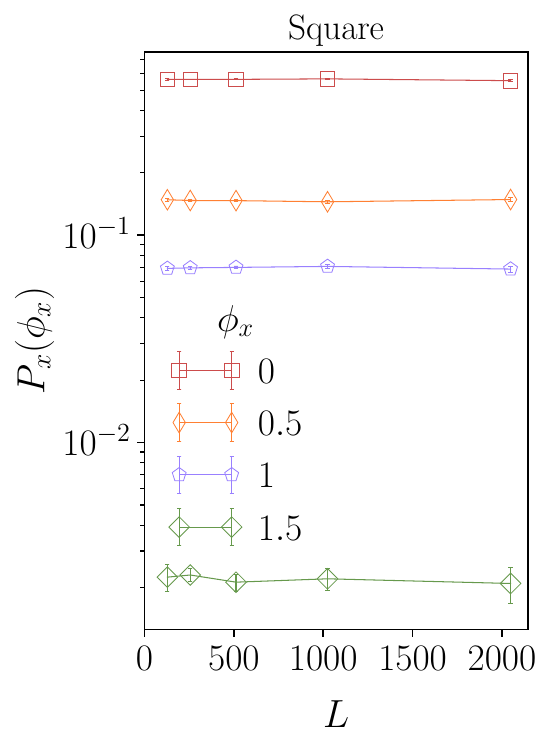}}
	\caption{\label{fig:Pxcriticalsquare} (a): Integer fluxes are governed by a Gaussian distribution at the critical point $w_c$ on the square lattice. This is illustrated by our data for the marginal distribution $P_x(\phi_x)$ restricted to integer values of $\phi_x$ at criticality. The corresponding critical stiffness is $g_c \approx 0.66(1)$.  (b): Fractional flux values occur at the critical point with probability that is $L$-independent in the thermodynamic limit. As is clear from the corresponding results in panel (a), which shows data for integer fluxes (labeled $Z$) and fractional fluxes (labeled $Z^{c}$) with different color coded symbols,  these probabilities for fractional fluxes are not governed by the same Gaussian distribution that controls the integer flux sectors.}
\end{figure}

The critical scaling of $\chi$ reflects the fact that the largest loop length at criticality scales as $s_{\rm max} ^{\rm crit} \sim L^{\theta_c}$ with $\theta_c \neq \theta_{6V}$. Indeed, we find that the critical loop size distribution also has a power-law exponent $\tau_c$ that is different from $\tau_{6V}$. This is evident from the scaling collapse of our data on both lattices for the critical loop size distribution (see Fig.~\ref{fig:critloopsize}):
\begin{eqnarray}
P_l^{\rm crit}(s,L) &=& \frac{c_7}{L^{\tau_{c} \theta_{c}}} G_{\rm crit}\left(\frac{c_8 s}{L^{\theta_{c}}}\right)
\label{eq:Pcrit}
\end{eqnarray}
with $\theta_c \approx 1.375(10)$ as before, and $\tau_c \approx 2.467(10)$; thus $\tau_c$ is also significantly different from the six vertex value $\tau_{6V} = 2.142857\dots$. For $x \ll 1$, $G_{\rm crit}(x) \sim x^{-\tau_{c}}$. And for $x \gg 1$, $G_{\rm crit}(x)$ vanishes rapidly with increasing $x$. Note that the measured values of $\theta_c$ and $\tau_c$ also satisfy the scaling relation $\tau_c \theta_c = \theta_c + 2$ within the errors of our measurement, suggesting that $\zeta_c = 0$, consistent with the observation that the ratio of the sizes of the second largest and largest loop scales to zero with increasing size.

The probability distribution $P(\phi_x, \phi_y)$ takes on a limiting Gaussian form at large $L$ in both phases, but fractional flux values are governed by this limiting Gaussian distribution only in the $w<w_c$ phase. In other words, for $w<w_c$ on both lattices, integer and fractional values of the fluxes are both governed by the same Gaussian distribution in the thermodynamic limit. In contrast, for $w>w_c$ on both lattices, this limiting Gaussian distribution only describes the sector with both $\phi_x$ and $\phi_y$ integers; indeed, $P(\phi_x, \phi_y) \rightarrow 0$ as $L \rightarrow \infty$ unless {\em both} $\phi_x$ and $\phi_y$ are integers.
This is illustrated for the square lattice in Fig.~\ref{fig:fluxsquareandhoneycomb} (top panel), and in Fig.~\ref{fig:fluxsquareandhoneycomb} (bottom panel) for the honeycomb lattice. Note that the square lattice distribution is actually a product of independent Gaussian distributions for $\phi_x$ and $\phi_y$: 
\begin{eqnarray}
P(\phi_x,\phi_y) & \propto & \exp(-\pi g (\phi_x^2+\phi_y^2))
\label{eq:Pjointsquare}
\end{eqnarray}
In contrast, on the honeycomb lattice, we find  
\begin{eqnarray}
P(\phi_x,\phi_y) \propto \exp(-\pi g (\phi_x^2+\phi_y^2 + \phi_x \phi_y)) \; ,
\label{eq:Pjointhoneycomb}
\end{eqnarray}
which couples $\phi_x$ and $\phi_y$ as expected on symmetry grounds~\cite{Patil_Dasgupta_Damle_2014}. \begin{figure}
	\centering
	\subfigure[]{\includegraphics[width=0.49\linewidth]{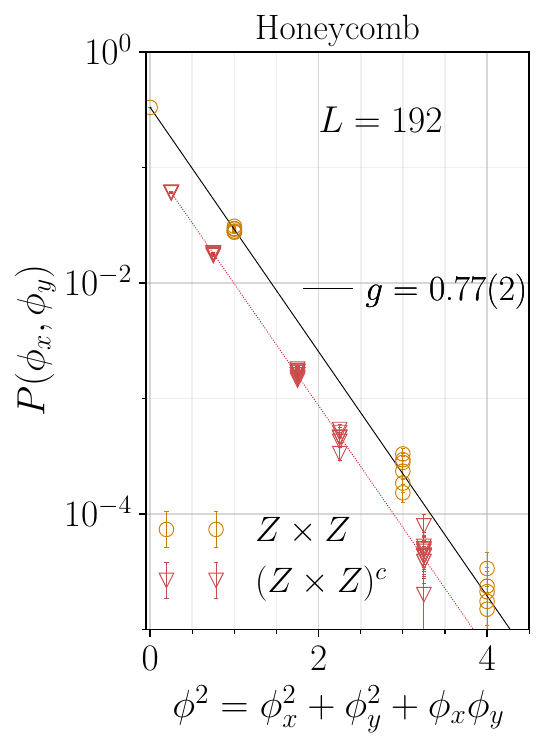}}
	\subfigure[]{\includegraphics[width=0.49\linewidth]{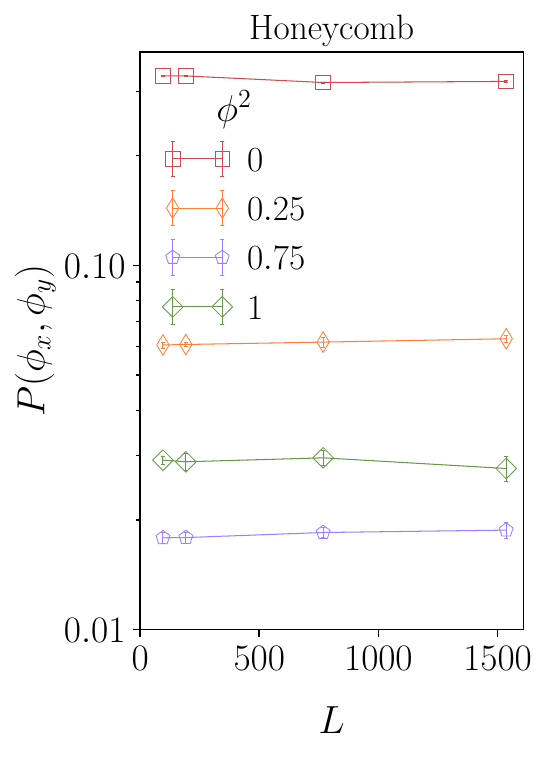}}
	\caption{\label{fig:Pjointcriticalhoneycomb} (a): Integer fluxes are governed by a Gaussian distribution at critical point $w_c$ on the honeycomb lattice. This is illustrated by our data for the joint distribution $P(\phi_x,\phi_y)$ at criticality. The corresponding stiffness at criticality is estimated to be $g_c \approx 0.77(2)$. (b): Fractional flux values occur at the critical point with probability that is $L$-independent in the thermodynamic limit. As is clear from the corresponding results in panel (a), which shows data for the sector with both $\phi_x$ and $\phi_y$ integer (labeled $Z\times Z$) and data for all other sectors (collectively labeled $(Z \times Z)^{c}$) with different color coded symbols, these probabilities for fractional fluxes are not governed by the same Gaussian distribution that controls the integer flux sectors.}
\end{figure}


$P_{\rm int}(\phi_x, \phi_y)$, the {\em restriction} of $P(\phi_x, \phi_y)$ to integer values of $\phi_x$ and $\phi_y$, has an $L$-independent Gaussian form in the thermodynamic limit when $w=w_c$, and the corresponding stiffness constant $g$ varies smoothly with $w$ across the transition. Interestingly, fractional values of flux also survive in the thermodynamic limit at the critical point, but are governed by a different $L$-independent limiting distribution.
 We illustrate this on the square lattice with our data for the marginal distribution $P_x(\phi_x)$ in Fig.~\ref{fig:Pxcriticalsquare}. For the honeycomb lattice case, this is illustrated in Fig.~\ref{fig:Pjointcriticalhoneycomb} which shows our results for the joint distribution $P(\phi_x, \phi_y)$ at criticality (in the honeycomb case, the symmetry of the underlying triangular Bravais lattice implies that there are two sectors: a sector in which $\phi_x$, $\phi_y$ and $\phi_x+\phi_y$ are all integers, and another sector in which two of these are fractional and one is an integer).  \begin{figure}
	\centering
	\subfigure[\label{}]{\includegraphics[width=0.49\linewidth]{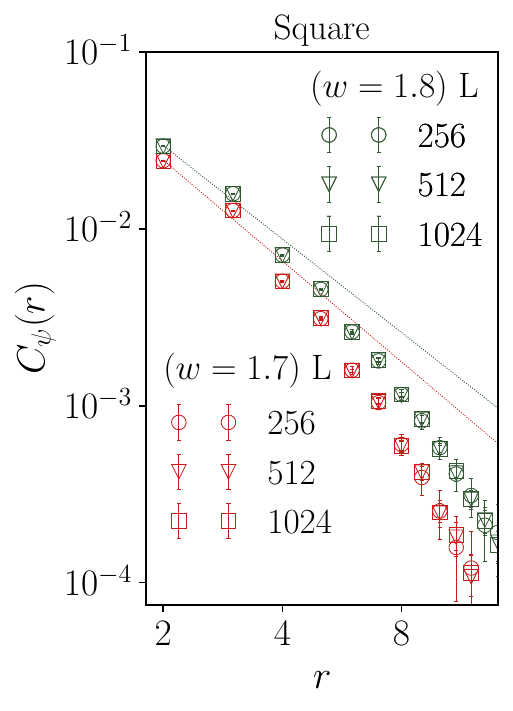}}
	\subfigure[\label{}]{\includegraphics[width=0.49\linewidth]{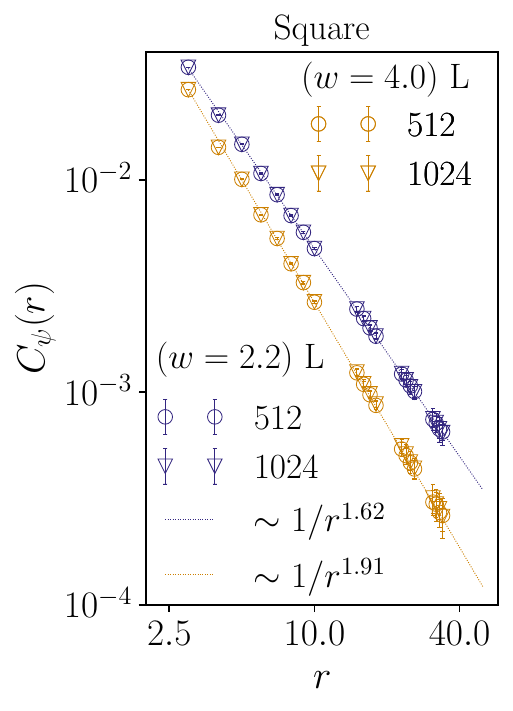}}
	\caption{\label{fig:squarecolumnarpowerlaw} (a) The two-point correlation function $C_\psi(r)$ of the columnar order parameter $\psi$ in the flux-fractionalized phase for $w<w_c$ on the square lattice decays rapidly to zero, faster than any power law. (b) In contrast, in the short-loop phase for $w>w_c$, it displays power-law behavior.}
\end{figure}

\subsection{Coarse-grained action and correlation functions away from $w=w_c$}
\label{subsec:offcriticaltheory}

We note that the Gaussian flux distributions at all $w \neq w_c$ in effect {\em constitute a measurement } of a Gaussian effective action for a coarse-grained height field $h(r)$ that describes the long-wavelength behavior of the dimer-loop model in either phase away from the critical point at $w=w_c$.
On the square lattice, this action has the form 
\begin{eqnarray}
S &= & \pi g \int d^2x (\partial_\mu h)^2 \; ,
\label{eq:Seffsquare}
\end{eqnarray} 
with $g$ obtained from the measured Gaussian flux distribution. On the honeycomb lattice, the corresponding effective action for $h(r)$ has symmetries of the underlying triangular Bravais lattice~\cite{Fradkin_Huse_Moessner_etal_2004}. These symmetries are most conveniently incorporated by working with a Gaussian effective action for real-valued $h(r)$ defined on sites $r$ of a coarse-grained triangular lattice~\cite{Patil_Dasgupta_Damle_2014}: 
\begin{eqnarray}
S &=& \frac{\pi g}{2} \sum_{\langle r r' \rangle} (h_r -h_{r'})^2 \; ,
\label{eq:Seffhoneycomb}
\end{eqnarray} 
 where $\langle r r' \rangle$ represents nearest-neighbor links of this triangular lattice, and $g$ is obtained from the measured Gaussian flux distribution. \begin{figure}
	\centering
	\subfigure[\label{}]{\includegraphics[width=0.49\linewidth]{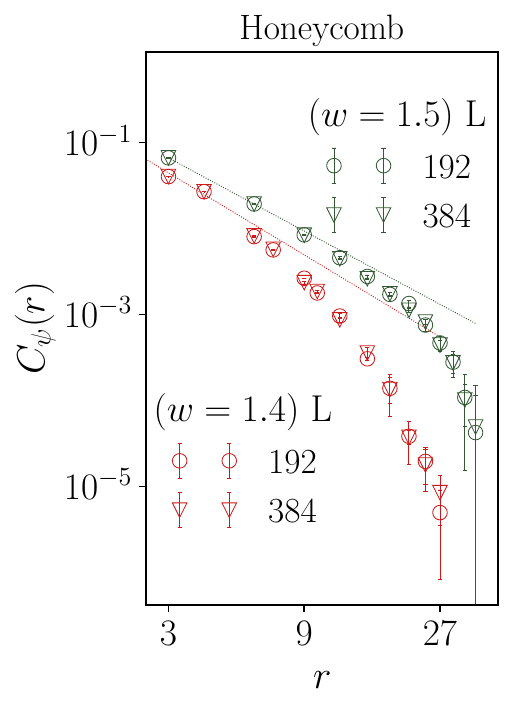}}
	\subfigure[\label{}]{\includegraphics[width=0.49\linewidth]{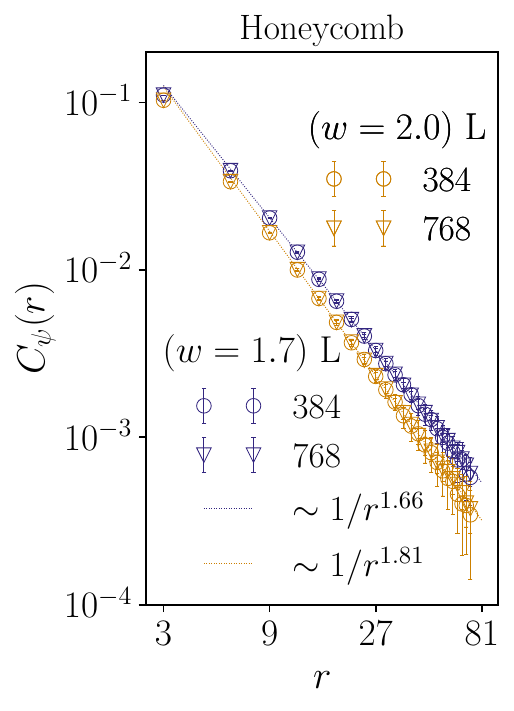}}
	\caption{\label{fig:honeycombcolumnarpowerlaw} (a) The two-point correlation function $C_\psi(r)$ of the columnar  order parameter $\psi$ for ordering at the three-sublattice wavevector on the honeycomb lattice decays rapidly (faster than a power law) to zero in the flux-fractionalized phase for $w<w_c$. (b) In contrast, in the short-loop phase for $w>w_c$, it displays power-law behavior.}
\end{figure}

For both lattices, these Gaussian effective field theories have the same form on either side of the critical point. However, the fact that fractional fluxes survive in the thermodynamic limit in one phase, but are excluded in the other phase at large $L$ implies that the {\em physical} operators in the field theory, {\em i.e.} operators whose correlators can potentially represent correlation functions of microscopic observables of the dimer-loop model, are constrained by different single-valuedness requirements on two sides of the transition: For $w>w_c$, all physical operators must be single-valued under $h(r) \rightarrow h(r)+1$, while for $w<w_c$, they must be single-valued under $h(r) \rightarrow h(r)+1/2$. Thus, in field-theoretical language~\cite{DiFrancesco_Mathieu_Senechal_book_1997}, this unusual transition at $w_c$ corresponds to a jump in the ``compactification radius'' of the free scalar field  $h$. This has interesting consequences for correlation functions of local observables. 

A good example is provided by the complex columnar order parameter field $\psi(r)$ on the square lattice, defined as in Ref.~\cite{Ramola_Damle_Dhar_2015} (see Sec.~\ref{sec:AlgoAndMeasurements}). The real part of $\psi$ measures the columnar ordering of dimer (and loop) occupation numbers of horizontal links at wavevector $(\pi, 0)$, while the imaginary part measures the columnar ordering  of dimer (and loop) occupation numbers of vertical links at wavevector $(0,\pi)$. The leading effective field theory operator whose transformation properties under square lattice symmetries match those of $\psi(r)$ is $\exp(2\pi i h(r))$~\cite{Fradkin_Huse_Moessner_etal_2004,Alet_Ikhlef_Jacobsen_etal_2006,Moessner_Tchernyshyov_Sondhi_2004,
Patil_Dasgupta_Damle_2014,Ramola_Damle_Dhar_2015}. This obeys the relevant single-valuedness criterion for $w>w_c$, but not for $w<w_c$.  \begin{figure} 
	\centering
	\subfigure[\label{fig:halfvortexsquarewless}]{\includegraphics[width=0.49\linewidth]{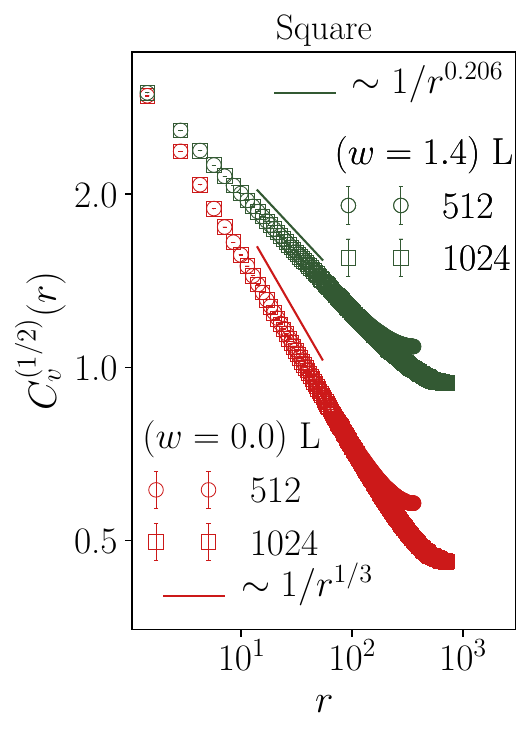}}
	\subfigure[\label{fig:halfvortexhoneycombwless}]{\includegraphics[width=0.49\linewidth]{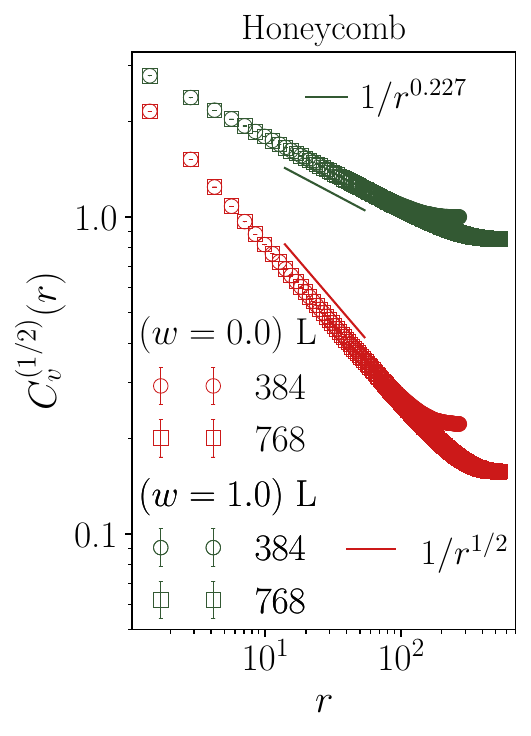}}
	\caption{\label{fig:halfvortexwless} The half-vortex correlation function shows power-law behavior for $w<w_c$ on (a) the square lattice, and (b) the honeycomb lattice.  }
\end{figure}

On the honeycomb lattice, the microscopic observable that corresponds to the vertex operator $\exp(2\pi i h(r))$ of the effective field theory is again the local order parameter field $\psi$ for columnar order at the three-sublattice ordering wavevector of the underlying triangular Bravais lattice~\cite{Fradkin_Huse_Moessner_etal_2004,Patil_Dasgupta_Damle_2014}.  For the correponding spin system on the kagome lattice, this is the local order parameter field for three-sublattice spin order at wavevector $(\vec{G}_1+\vec{G}_2)/6$, where $\vec{G}_x = (8\pi/b \sqrt{3}) \times \hat{k}_x$ and $\vec{G}_y = (8\pi/b \sqrt{3}) \times \hat{k}_y$ and $b$ is the lattice periodicity of the underlying triangular Bravais lattice (see Fig.~\ref{fig:spinlattice} for the orientation of the unit vectors $\hat{k}_x$ and $\hat{k}_y$ and the definition of $b$).

Therefore, on the square lattice, we expect that the $w>w_c$ phase has power-law columnar order with exponent $1/g$, as predicted by the fact that $\langle e^{-2\pi i h(r)} e^{+2\pi i h(0)} \rangle \sim 1/r^{1/g}$ in the effective field theory with action given by Eq.~\ref{eq:Seffsquare}.  On the honeycomb lattice, we again expect power-law columnar order in the $w>w_c$ short loop phase. The corresponding exponent is expected to be $2/g\sqrt{3}$ in this case, as predicted by the fact that $\langle e^{-2\pi i h(r)} e^{+2\pi i h(0)} \rangle \sim 1/r^{2/g\sqrt{3}}$ when the effective action takes on the form given in Eq.~\ref{eq:Seffhoneycomb}. In sharp contrast, the $w<w_c$ flux-fractionalized phase is expected to have rapidly decaying short-ranged correlations of $\psi(r)$ on both lattices. Our numerical results on either side of the transition are consistent with this prediction on both lattices. This is illustrated in Fig.~\ref{fig:squarecolumnarpowerlaw} and in Fig.~\ref{fig:honeycombcolumnarpowerlaw}. \begin{figure} 
	\centering
	\subfigure[\label{fig:halfvortexsquarewgreat}]{\includegraphics[width=0.49\linewidth]{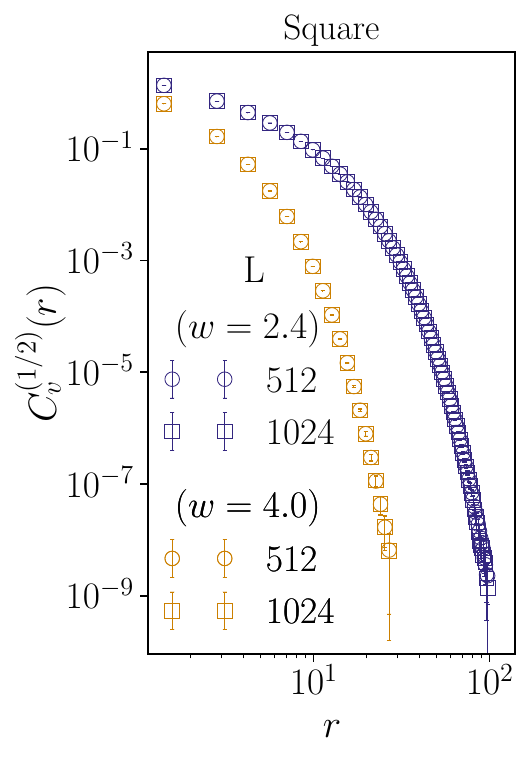}}
	\subfigure[\label{fig:halfvortexhoneycombwgreat}]{\includegraphics[width=0.49\linewidth]{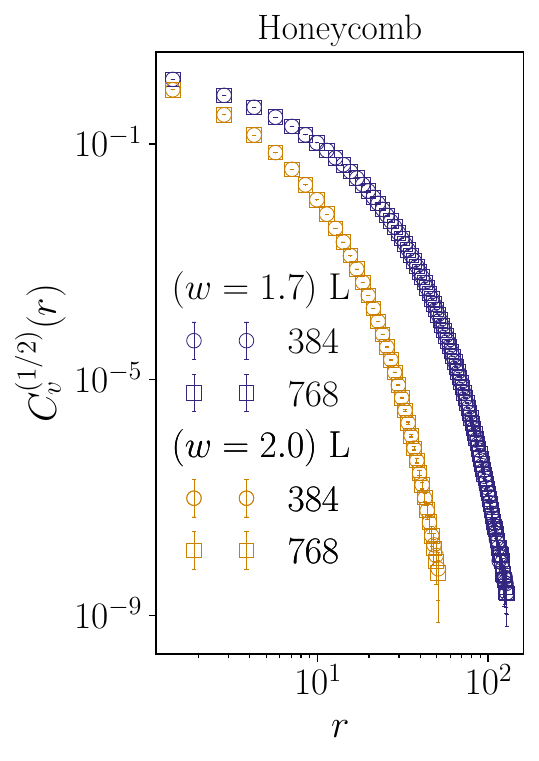}}
	\caption{\label{fig:halfvortexwgreat} The half-vortex correlation function in the $w>w_c$ short-loop phase decays rapidly, faster than any power-law, on (a) the square lattice, and (b) the honeycomb lattice.}
\end{figure}

Indeed, we see from this argument that the leading allowed vertex operator in the flux-fractionalized phase is $\exp(4\pi i h(r))$. On the square lattice, this represents the nematic order parameter for loop and dimer occupation variables~\cite{Moessner_Tchernyshyov_Sondhi_2004,Ramola_Damle_Dhar_2015}. Since $\langle e^{-4\pi i h(r)} e^{+4\pi i h(0)} \rangle \sim 1/r^{4/g}$ in the square lattice case, this implies that the local nematic order parameter in the square latticec case has power-law correlations $\sim 1/r^{4/g}$. From the measured value of $g$ (obtained from fits to $P(\phi_x,\phi_y)$), we find (as we discuss later in this section) that $4/g > 2$ throughout the flux-fractionalized phase. On the honeycomb lattice, we have $\langle e^{-4\pi i h(r)} e^{+4\pi i h(0)} \rangle \sim 1/r^{8/g\sqrt{3}}$. From the measured values of $g$, we find that this exponent satisfies the analogous inequality $8/g\sqrt{3} >2$ throughout the flux-fractionalized long loop phase. Therefore, the correlation functions of dimers and loop segments in the flux-fractionalized phase on both lattices are expected to be dominated by their dipolar parts that are represented in the effective field theory by correlation functions of gradients of $h(r)$; these dipolar contributions have a characteristic $1/r^2$ power-law behavior (modulated in the square lattice case by a sublattice-dependent sign)~\cite{Youngblood_Axe_McCoy_1980,Youngblood_Axe_1981,Fradkin_Huse_Moessner_etal_2004,
Moessner_Tchernyshyov_Sondhi_2004,Henley_2010}.

In sharp contrast, the corresponding correlations in the $w>w_c$ short-loop phase are dominated by the power-law columnar order that characterises this phase, since the corresponding power-law exponent is found to be consistently smaller than $2$ on both lattices; an illustration of this was already provided in Fig.~\ref{fig:squarecolumnarpowerlaw} and Fig.~\ref{fig:honeycombcolumnarpowerlaw}. This is consistent  with the the values of $g$ obtained from fits to $P(\phi_x, \phi_y)$ in this phase, since the field-theoretical prediction for the power-law exponent for columnar order is $1/g$ ($2/g\sqrt{3}$) on the square (honeycomb) lattice, and the values of $g$ obtained from $P(\phi_x,\phi_y)$ satisfy $g>1/2$ ($g > 1/\sqrt{3}$) throughout the $w>w_c$ short-loop phase on the square (honeycomb) lattice for any finite $w$ accessible to our numerics.

Another illustration is provided by the correlation function of a pair of oppositely charged test vortices with charge $\pm 1/2$, which, at a formal level,  can be unambiguously computed from these Gaussian field theories on either side of the transition. On the square lattice, we obtain $C^{(q)}_v(r) \sim 1/r^{q^2g}$ for the correlation function of a vortex-antivortex pair with vorticity $\pm q$. On the honeycomb lattice, we obtain $C^{(q)}_v(r) \sim 1/r^{q^2g\sqrt{3}/2}$. We compare these predictions for $q=1/2$ with numerical results for  $C^{(1/2)}_v(r)$ obtained from the histogram of the head-to-tail displacements in the half-vortex worm update.
 For the half-vortex correlator, we find that these predictions provide an accurate fit to our data {\em only} for $w<w_c$.  For $w>w_c$, the half-vortex correlation function decays rapidly to zero. This is shown in Fig.~\ref{fig:halfvortexwless} and Fig.~\ref{fig:halfvortexwgreat}. We understand these contrasting behaviors to be a  direct consequence of the fact that fractional fluxes are allowed for $w<w_c$ but excluded from the system in the thermodynamic limit when $w>w_c$. \begin{figure} 
	\centering
	\subfigure[\label{fig:unitvortexsquarewless}]{\includegraphics[width=0.49\linewidth]{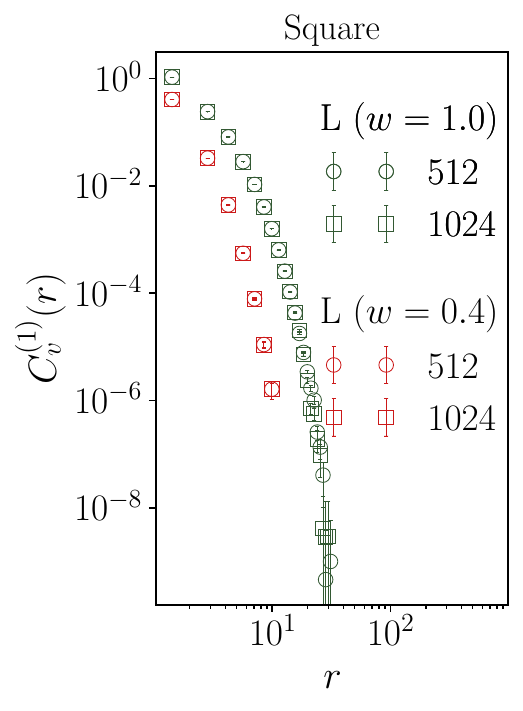}}
	\subfigure[\label{fig:unitvortexhoneycombwless}]{\includegraphics[width=0.49\linewidth]{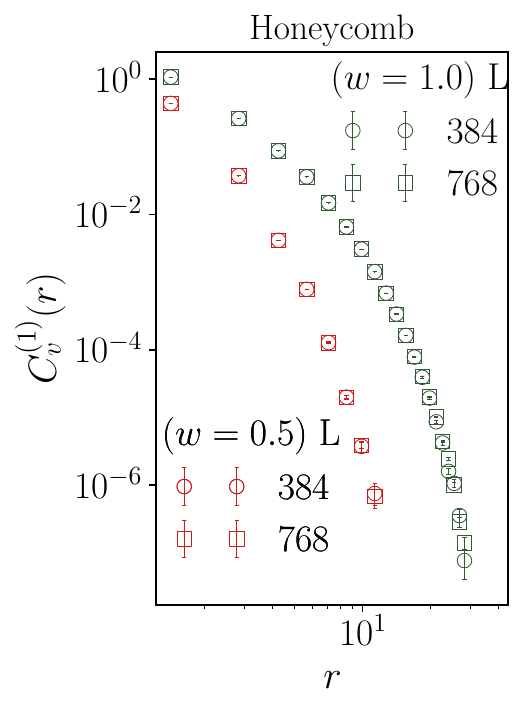}}
	\caption{\label{fig:unitvortexwless}  The unit-vortex correlation function in the $w<w_c$ flux-fractionalized phase decays rapidly, faster than any power-law, on (a) the square lattice, and (b) the honeycomb lattice.}
\end{figure}

\begin{figure} 
	\centering
	\subfigure[\label{fig:unitvortexsquarewgreat}]{\includegraphics[width=0.49\linewidth]{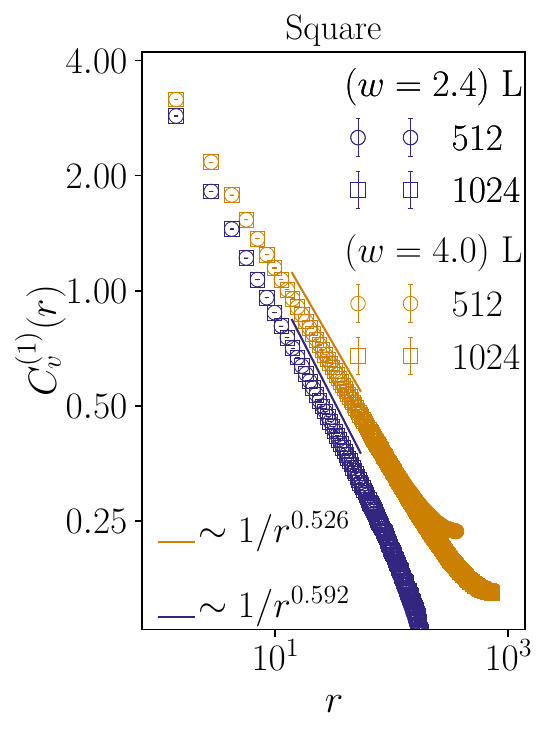}}
	\subfigure[\label{fig:unitvortexhoneycombwgreat}]{\includegraphics[width=0.49\linewidth]{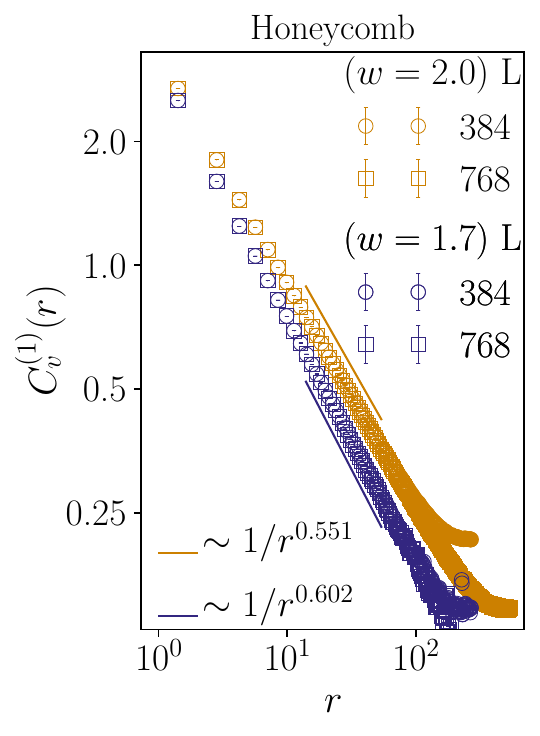}}
	\caption{\label{fig:unitvortexwgreat} The unit-vortex correlation function shows power-law behavior for $w>w_c$ on (a) the square lattice, and (b) the honeycomb lattice.}
\end{figure}

We also find that the corresponding effective field theory predictions (now with $q=1$) for unit-vortex correlation functions $C^{(1)}_v(r)$ provide an accurate fit to the observed behavior {\em only} when $w> w_c$. When $w<w_c$, the unit-vortex correlation function decays rapidly to zero with increasing $r$. This behavior of unit-vortex correlation function on both lattices for $w<w_c$ $(w>w_c)$ is shown in  Fig.~\ref{fig:unitvortexwless} (Fig.~\ref{fig:unitvortexwgreat}). Unlike the $w>w_c$ behavior of the half-vortex correlators, this observation about unit-vortex correlators in the $w<w_c$ flux-fractionalized phase does not follow immediately from our earlier result on flux sectors that survive in the thermodynamic limit, and we do not have a complete theory for this behavior. 

Nevertheless, this striking behavior of the unit vortex correlator, when viewed in conjunction with the earlier results on the half-vortex correlator, lead to the following heuristic picture: In the
flux-fractionalized phase, a pair of oppositely charged test vortices with vorticity $\pm 1/2$ feel a logarithmic attraction of entropic origin, similar to the logarithmic attraction felt by a test pair of oppositely charged unit-vortices in the $w > w_c$ phase. On the other hand, when $w > w_c$ , half-vortices are bound
into unit-vorticity pairs and cannot move by themselves over large scales. Conversely, in the $w < w_c$ phase, unit-vortices are unstable to breaking up into two half-vortices. 

\begin{figure}
	\centering
	\subfigure[\label{}]{\includegraphics[width=0.49\linewidth]{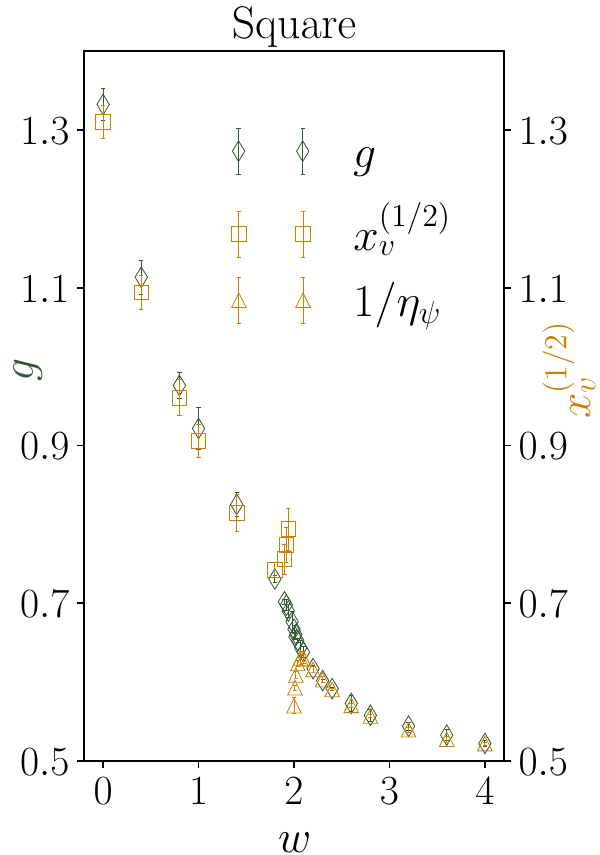}}
	\subfigure[\label{}]{\includegraphics[width=0.49\linewidth]{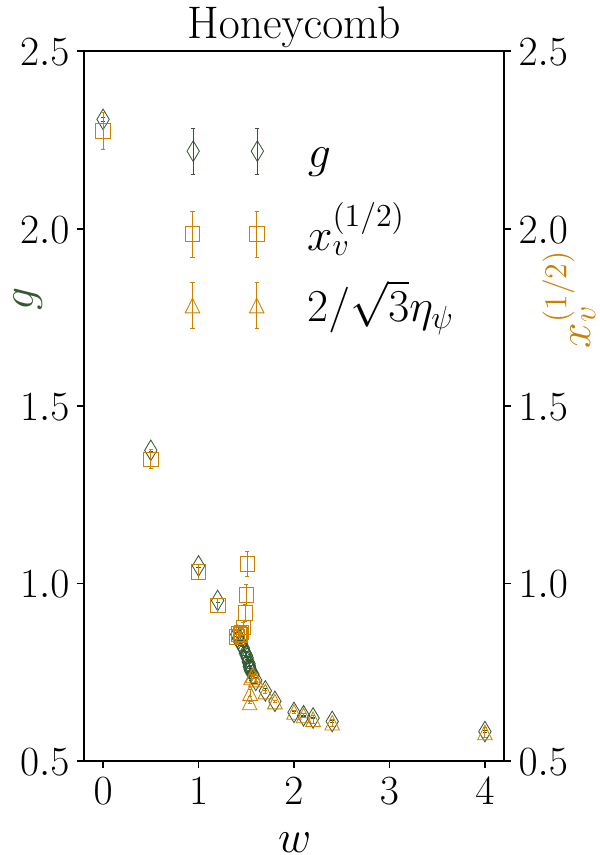}}
	\caption{\label{fig:gvsw} The value of the stiffness $g$ that enters the effective field theory can be extracted from fits to the flux probability distribution $P(\phi_x, \phi_y)$ for all values of $w$. At the critical point $w=w_c$, and for $w>w_c$, this can be done by studying the restriction $P_{\rm int}$ to integer values of $\phi_x$ and $\phi_y$, while all values of flux can be used to perform the fits in the flux-fractionalized phase for $w<w_c$. The resulting $g(w)$ obtained in this way is shown for the square lattice in (a) and the honeycomb lattice in (b). The same stiffness $g$ can also be extracted by fitting the power-law decay of the columnar order parameter correlations in the $w>w_c$ short-loop phase. This is illustrated in (a) for the square lattice; in the vicinity of the critical point, this estimate deviates significantly from $g$ obtained using fits to $P(\phi_x,\phi_y)$. In addition, one can also obtain $g$ by fits to the power-law behavior of the half-vortex correlator in the $w<w_c$ flux-fractionalized phase. This is illustrated in (a) for the square lattice and (b) for the honeycomb lattice; deviations from $g$ obtained using fits to $P(\phi_x,\phi_y)$ are again visible in the vicinity of the critical point. For the square (honeycomb) lattice, $x_v^{(1/2)} \equiv 4\eta_v^{(1/2)}$ ($x_v^{(1/2)} \equiv 8\eta_v^{(1/2)}/\sqrt{3}$).}
\end{figure}

Returning to a more quantitative analysis, we note that we can obtain an independent measurement of $g(w)$ by fitting the observed power-law behaviors of half-vortex correlators to the field-theoretical predictions for $w< w_c$. For $w>w_c$, we can obtain an independent measurement of $g(w)$ from the power-law behavior of the columnar order parameter correlation function $C_{\psi}$ and from the power-law behavior of the unit-vortex correlators. We find that the numerical values of $g(w)$ obtained in this way are consistent within our errors with those obtained from fits to the flux distribution function $P(\phi_x, \phi_y)$ in both phases. This is shown in Fig.~\ref{fig:gvsw}. Thus, our results pass this important consistency check on the validity of such an effective-field theory for the dimer-loop model in both phases away from the critical point at $w=w_c$.

Additionally, the measured values for the honeycomb lattice also satisfy $g(w=0)= 4 \times g(w=\infty) = 4/\sqrt{3}$, consistent with the known value for the fully-packed dimer model and the mapping between $Z(w=0)$ and $Z(w = \infty)$ alluded to earlier. Moreover, the measured value of $g(w=0)$ on the square lattice agrees within numerical error with the theoretical expectation that $g(w=0)=g_{6V} \equiv 4/3$ for the six-vertex model~\cite{diFrancesco_Saleur_Zuber_1987}. 

However, close to the critical point on both lattices, we do find that the power-law exponents for the half-vortex correlator in the flux-fractionalized phase and the unit-vortex correlator in the short-loop phase deviate from the predictions of the effective field theory (using the value of $g$ obtained from fits to $P(\phi_x,\phi_y)$); this is also visible in the estimates of $g(w)$ displayed in Fig.~\ref{fig:gvsw}. We ascribe this to the proximity of the critical point at $w=w_c$, which cannot be described by such a Gaussian effective action. This is explored further in Sec.~\ref{subsec:criticaltheory}.

\subsection{Comments on the critical theory at $w=w_c$}
\label{subsec:criticaltheory}
Away from $w=w_c$, in both phases, all the flux sectors that survive in the thermodynamic limit are governed by a Gaussian distribution. Therefore, as we have discussed in Sec.~\ref{subsec:offcriticaltheory}, one has a consistent description in terms of a Gaussian effective action for a coarse-grained height field $h$, augmented by an appropriate choice of compactification radius that restricts the set of physical operators of the theory.  \begin{figure} 
	\centering
	\subfigure[\label{fig:halfvortexsquarewcrit}]{\includegraphics[width=0.49\linewidth]{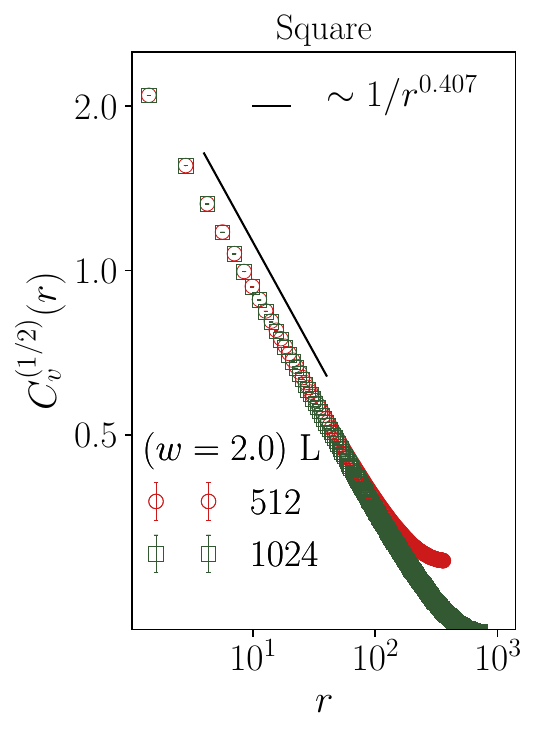}}
	\subfigure[\label{fig:halfvortexhoneycombwcrit}]{\includegraphics[width=0.49\linewidth]{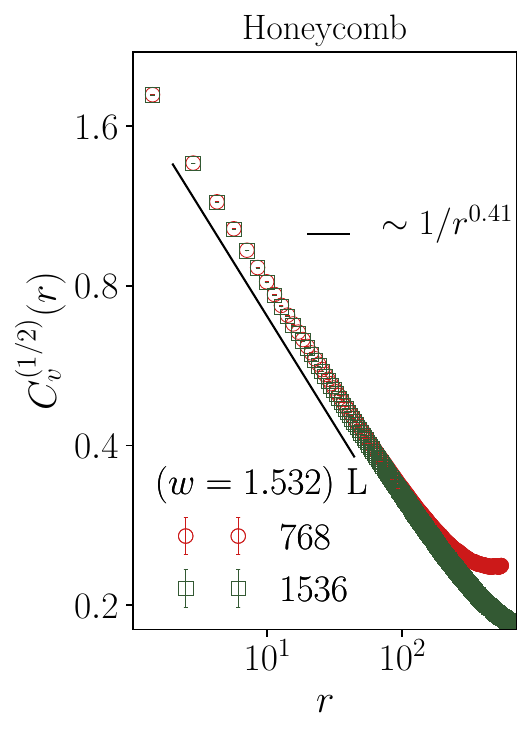}}
	\caption{\label{fig:halfvortexwcrit} (a) and (b): The critical half-vortex correlation function decays as a power-law, with power-law exponent $\eta_v^{(1/2)}(w_c) \approx 0.409(5)$ on both the square and the honeycomb lattice.}
\end{figure}

The picture is quite different at the $w=w_c$ critical point.
As is evident from the data already analyzed in Fig.~\ref{fig:Pxcriticalsquare} and Fig.~\ref{fig:Pjointcriticalhoneycomb}, fractional fluxes also survive in the thermodynamic limit at $w=w_c$ in addition to integer fluxes,  but they are not governed by the $L$-independent Gaussian distribution that controls the integer fluxes. As a result, one cannot expect to describe or intepret long-wavelength properties at the critical point in terms of a free field theory for a coarse-grained height field in the same straightforward way that is successful in either phase.

\begin{figure} 
	\centering
	\subfigure[[\label{fig:criticalcolumnarpowerlawsqr}]{\includegraphics[width=0.49\columnwidth]{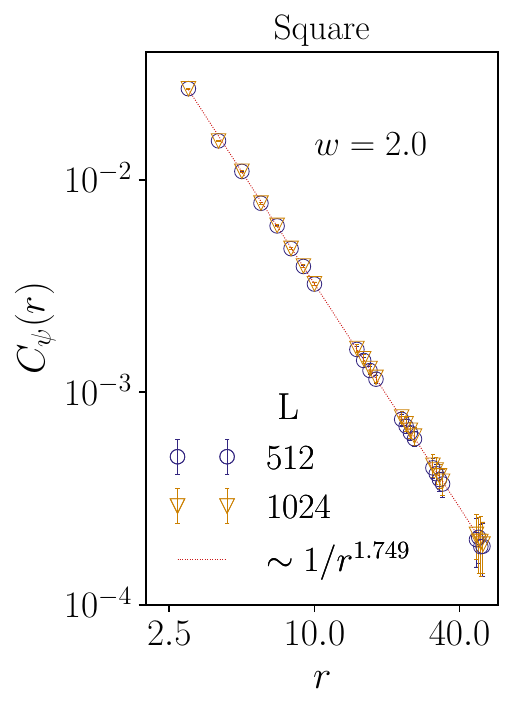}}
	\subfigure[[\label{fig:criticalcolumnarpowerlawhl}]{\includegraphics[width=0.49\columnwidth]{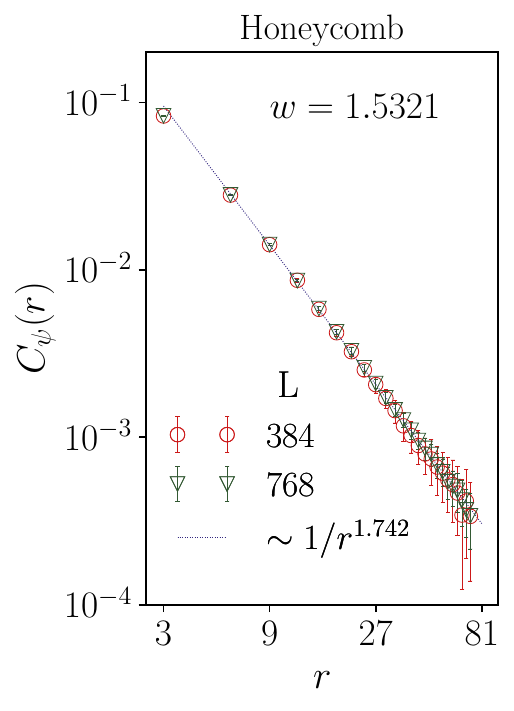}}
	\caption{\label{fig:criticalcolumnarpowerlaw}  (a) and (b): The columnar order parameter correlation function at criticality on both lattices decays as a power-law, with an exponent that takes on a universal value estimated to be $\eta_{\psi}(w_c) \approx 1.745(15)$.}
\end{figure}

\begin{figure} 
	\centering
	\subfigure[\label{fig:unitvortexsquarewcrit}]{\includegraphics[width=0.49\linewidth]{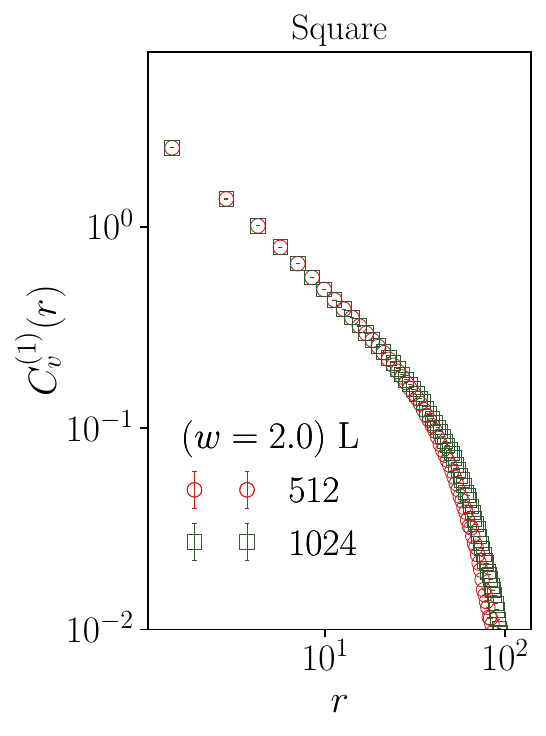}}
	\subfigure[\label{fig:unitvortexhoneycombwcrit}]{\includegraphics[width=0.49\linewidth]{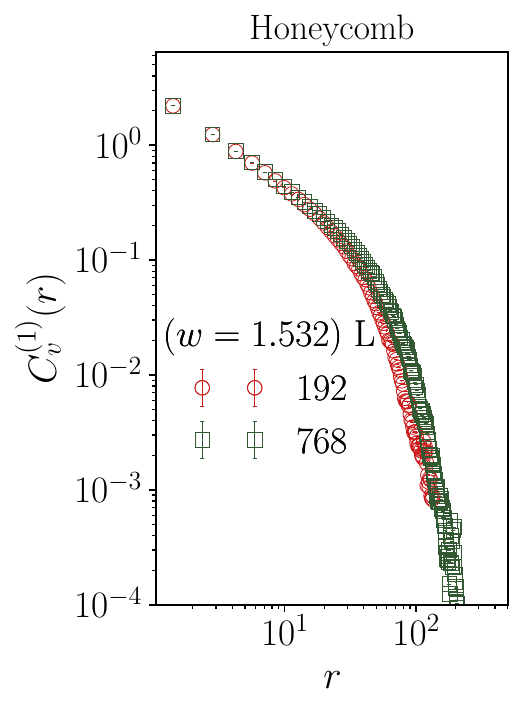}}
	\caption{\label{fig:unitvortexwcrit} (a) and (b): The unit-vortex correlation function at criticality decays rapidly, faster than any power law, on both the square and the honeycomb lattice.}
\end{figure}

To characterise the properties of the critical point at $w=w_c$, we measure the vortex correlation functions and the correlator of the columnar order parameter field $\psi$. We find that the half-vortex correlator has power-law behavior at the critical point both on the honeycomb lattice and on the square lattice, with an associated universal power-law exponent $\eta_{v}^{(1/2)}(w_c) \approx 0.409(5) $. This is displayed in Fig.~\ref{fig:halfvortexwcrit}. In addition, we find that the columnar order  parameter correlation function also has power-law behavior at criticality on both lattices, with the corresponding exponent being $\eta_{\psi}(w_c) \approx 1.75(2)$. This is shown in Fig.~\ref{fig:criticalcolumnarpowerlaw}.  However, the critical unit-vortex correlator on both lattices has a rapid fall off, faster than any power-law decay. This is clear from Fig.~\ref{fig:unitvortexwcrit}. 

The stiffness constants $g_c$ associated with the Gaussian distribution of integer fluxes at the critical point of the square and honeycomb lattices has been estimated in the fits already displayed in Fig.~\ref{fig:Pxcriticalsquare} and Fig.~\ref{fig:Pjointcriticalhoneycomb} respectively. From these fits, we find $g_c \approx 0.66(1)$ and $g_c \approx 0.77(2)$ for the square and the honeycomb lattice respectively.  
The values of the exponents $\eta_v^{(1/2)}(w_c)$ and $\eta_{\psi}(w_c)$ quoted earlier are clearly not consistent with the exponents one would predict for a Gaussian theory with stiffness given by these measured values of $g_c$ on both lattices. In addition, the short-ranged form of the unit-vortex correlator at criticality is also not consistent with predictions from a Gaussian effective action.
Thus, as already anticipated from our study of the critical flux distributions, it is clear that this combination of critical properties is not consistent with the predictions of a Gaussian effective action.  
However, we note that our measured values of $g_c$ on the two lattices do satisfy $g_c^{\rm square} = \sqrt{3} g_c^{\rm honeycomb}/2$ within the numerical errors associated with our fits.

In addition to correctly predicting the correlation length exponent $\nu$, any field-theoretical framework for understanding this critical point would have to correctly predict the values of the power-law exponents $\eta_{\psi}(w_c) \approx 1.75(2)$, $\eta_{v}^{(1/2)}(w_c) \approx 0.409(5)$, $\theta_c \approx 1.375(10)$, and $\tau_c \approx 2.47(1)$ (with the last two obeying $\tau_c \theta_c =  \theta_c+2$ within our numerical errors), associated respectively with the critical columnar order parameter correlation function, the critical half-vortex correlation function, and the critical loop size distribution. Identifying the correct critical theory remains a challenge, and we hope our fairly accurate numerical estimates of these exponents will be of some value in this regard.   

\subsection{Spin structure factor}
\label{subsec:spinstructurefactor}
From the point of view of potential experimental realizations of this physics in a kagome magnet, it is interesting to study the spin structure factor that can be probed by neutron scattering. For scattering of spin polarized neutrons in the ``classical'' temperature regime studied here, with the spin polarization being along the $z$ axis perpendicular to the kagome plane, the static non-spin-flip component of the scattering crosssection  is expected to directly probe the structure factor 
\begin{eqnarray}
S(q) &=& \langle S^z(q) S^z(-q)\rangle \; .
\label{eq:spinstructurefactor}
\end{eqnarray}
Here, the angular brackets denote the statistical average in our classical model, and $S^z(q)$ is the Fourier transform of the spins in one unit cell of the kagome lattice:
\begin{eqnarray}
 S^z(q) & = & \nonumber \\
 && \!\!\!\! \!\!\!\! \!\!\!\! \!\!\!\! \!\!\!\! \!\!\!\! \!\!\!\! \frac{1}{L} \sum_{R} e^{iq \cdot R} \left(S_1^z(R)e^{i q \cdot r_1} +S_2^z(R)e^{i q \cdot r_2} + S_3^z(R)e^{i q \cdot r_3} \right) \; , \nonumber \\
 &&
\end{eqnarray}
where the sum is over all up-pointing triangles of an $L \times L$ kagome lattice with $3L^2$ sites
and periodic boundary conditions, and $R+r_1$, $R+r_2$, and $R+r_3$ are coordinates of the three spins that belong to the up-pointing triangle whose center is at $R$. 

\begin{figure}
\includegraphics[width=\linewidth]{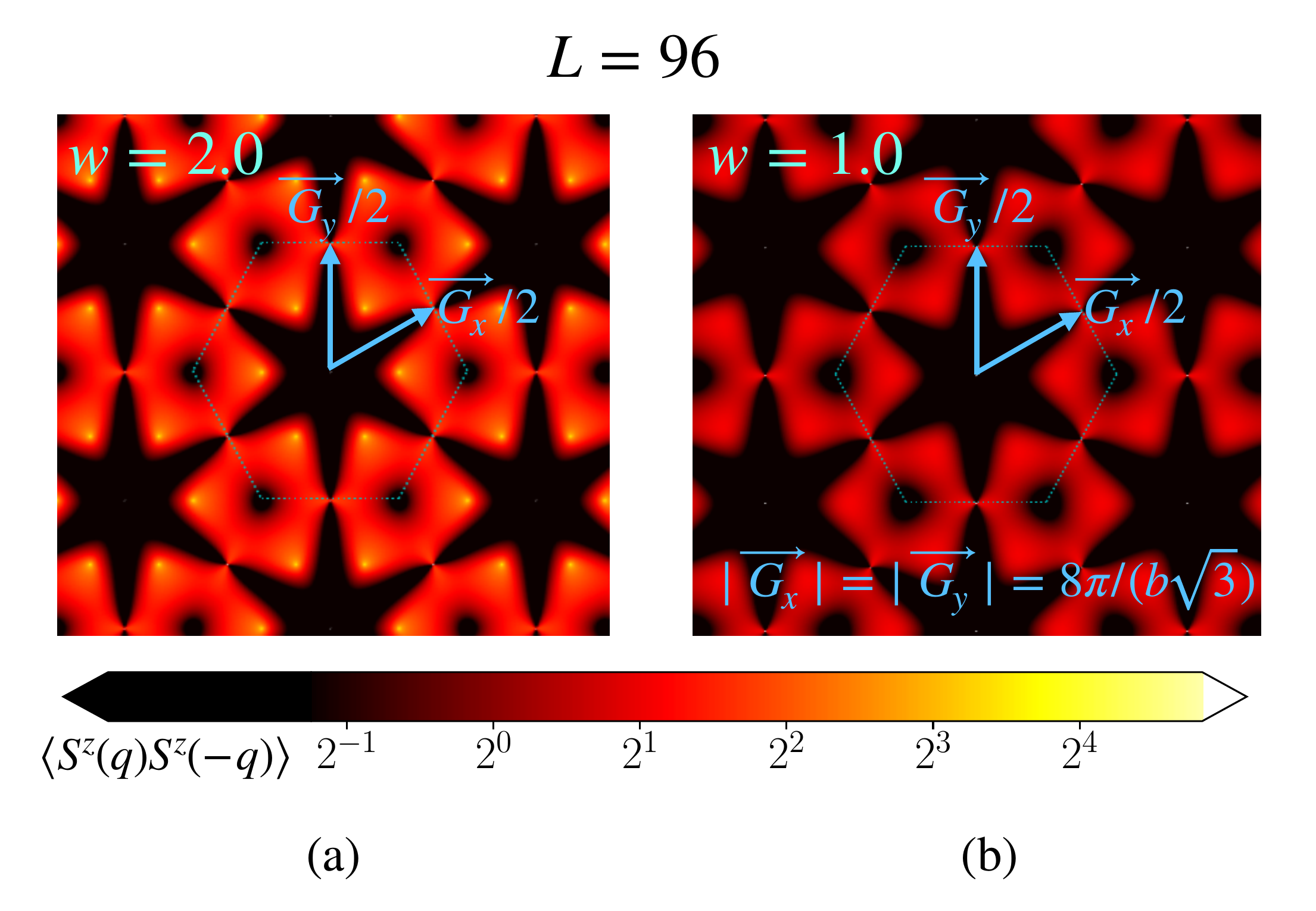}
\caption{Our numerical results  demonstrate that the kagome spin structure factor defined in Eq.~\ref{eq:spinstructurefactor} is sensitive to the flux-fractionalization transition at $w=w_c$. In (a), we see that the spin structure factor shows clearly visible peaks at wavevector $(\vec{G}_x+\vec{G}_y)/6$ and symmetry-related wavevectors, corresponding to power-law columnar order of spins in the $w>w_c$ phase, in addition to pinch-point singularities at the centers of the zone boundaries ({\em i.e.} at  wavevectors $\vec{G}_x/2$ and $\vec{G}_y/2$ and symmetry-related wavevectors). In b) we see that the same pinch-point singularities are visible in the $w<w_c$ long loop phase, but the peaks corresponding to power-law columnar order are absent. Here, $\vec{G}_x = (8\pi/b\sqrt{3}) \times \hat{k}_x$, $\vec{G}_y=(8\pi/b\sqrt{3}) \times \hat{k}_y$, and the unit vectors $\hat{k}_x$, $\hat{k}_y$ and the lattice constant $b$ are defined as in Fig.~\ref{fig:spinlattice}.}
\label{fig:structurefactor}
\end{figure}

With this motivation, we have measured $S(q)$ in both phases on either side of the flux fractionalization transition. For $w< w_c$, {\em i.e.} in the long loop phase, we find that the data exhibits a characteristic pinch-point structure at the centers of the zone boundaries, {\em i.e.} at  wavevectors $\vec{G}_x/2$ and $\vec{G}_y/2$ and symmetry-related wavevectors; here $\vec{G}_x = (8\pi/b \sqrt{3}) \times \hat{k}_x$ and $\vec{G}_y = (8\pi/b \sqrt{3}) \times \hat{k}_y$ and $b$ is the lattice periodicity of the underlying triangular Bravais lattice (see Fig.~\ref{fig:spinlattice} for the orientation of the unit vectors $\hat{k}_x$ and $\hat{k}_y$ and the definition of $b$). For $w>w_c$, {\em i.e.} in the short loop phase, the same pinch-point singularity continues to be clearly visible. In addition, there is a clearly identifiable peak at the columnar ordering wavevectors associated with three-sublattice order ($(\vec{G}_x+\vec{G}_y)/6$ and symmetry-related wavevectors), which distinguishes the $w>w_c$ power-law ordered short loop phase from the $w<w_c$ long loop phase. This is displayed in Fig.~\ref{fig:structurefactor}.
Thus, such a neutron measurement is sensitive to the power-law columnar order and its destruction at the flux fractionalization transition.

\subsection{Sanity check: Restriction to zero-flux sector}
\label{subsec:SanityCheck}
The theoretical arguments and computational results presented in the foregoing lead to a compelling picture, whereby power-law columnar order (three-sublattice order on the honeycomb lattice), characteristic of the $w>w_c$ short-loop phase, is destroyed in the $w<w_c$ long-loop phase by the proliferation of fractional fluxes.
However, fluxes or winding numbers are inherently boundary condition dependent concepts, while the presence or absence of power-law columnar order ought to be independent of boundary conditions. 

With this in mind, we  now restrict our computations to the zero-flux sector, and ask if our data restricted to this sector shows equally clear evidence of the same transition between a short-loop phase with power-law columnar order and a long-loop phase without it. To do this, we do not need to change the Monte Carlo algorithm or the periodic boundary conditions. Instead, we simply restrict ourselves to measuring physical observables only when the configuration belongs to the zero-flux sector. This is a valid (if slightly inefficient) procedure since it gives the correct relative weights to all configurations in the zero-flux sector.

\begin{figure}
	\centering
	\includegraphics[width=\linewidth]{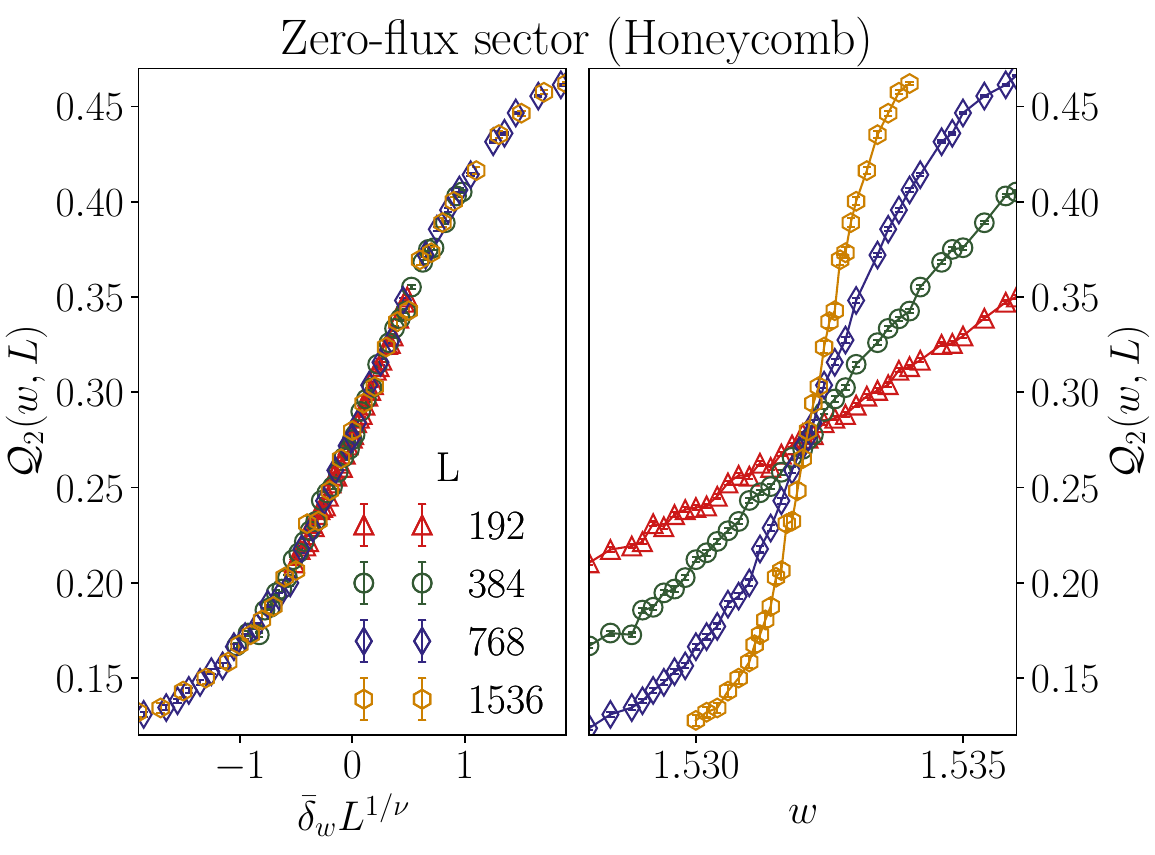}
	\caption{\label{fig:Q2zerowinding} Right panel: The Binder ratio ${\mathcal Q}_2$ of loop sizes (defined in Sec.~\ref{sec:AlgoAndMeasurements}) obtained from the zero-flux sector on the honeycomb lattice shows a clear crossing at a critical value $w_c\approx 1.5321(3)$. Left panel: For $w$ close to $w_c$, this zero-flux sector data for various sizes $L$ collapses on to the scaling form described in Eq.~\ref{eq:Q2AndPScaling}. The scaling collapse displayed here employs the following parameter values: $w_c = 1.5321$, $\nu=1.0$ and  $c_1=1$. }
\end{figure}

In Fig.~\ref{fig:Q2zerowinding}, we display the results of such a study of the Binder ratio ${\mathcal Q}_2$ (defined in Sec.~\ref{sec:AlgoAndMeasurements}) on the honeycomb lattice, now restricted to the zero-flux sector. Clearly, there is a transition from a short-loop phase to a long-loop phase at a value of $w_c$ that is, within errors, the same as our earlier unrestricted estimate. Further, we see that the data obeys a scaling collapse in the vicinity of the critical point, with a correlation length exponent that again matches within errors our earlier unrestricted estimate.

\begin{figure}
	\centering
	\includegraphics[width=\linewidth]{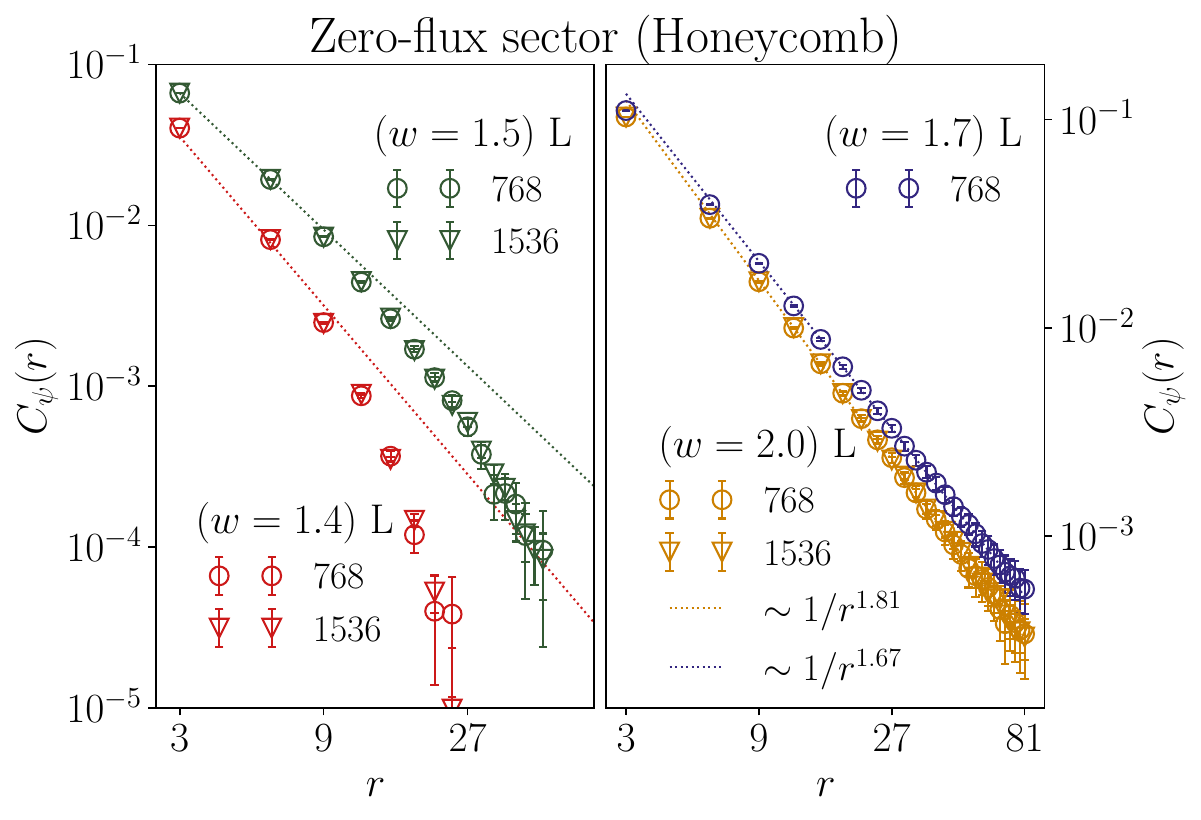}
	\caption{\label{fig:Cpsizerowinding} Left panel: The two-point correlation function $C_\psi(r)$ of the columnar  order parameter $\psi$ for ordering at the three-sublattice wavevector on the honeycomb lattice measured in the zero-flux sector of the configuration space in the flux-fractionalized phase ($w<w_c$) decays rapidly to zero, faster than any power law. Right panel: In contrast, in the short-loop phase for $w>w_c$, it displays power-law behavior. The power-law exponents for different $w$ obtained from this zero-flux sector data match within error with our previous unrestricted estimates shown in Fig.~\ref{fig:honeycombcolumnarpowerlaw}. }
\end{figure}

\begin{figure}
	\centering
	\includegraphics[width=\linewidth]{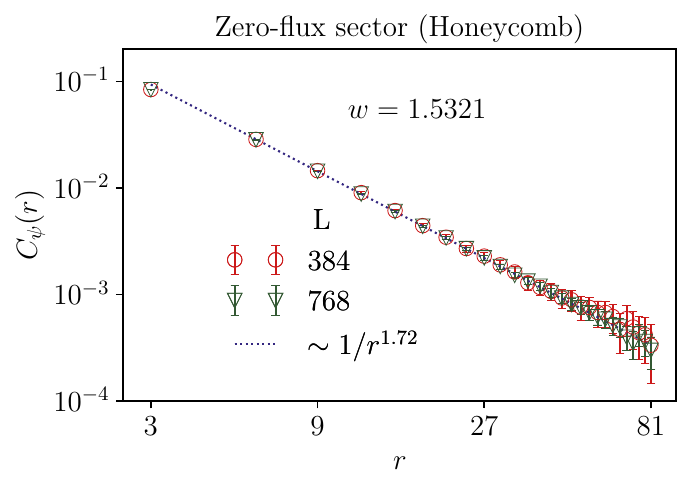}
	\caption{\label{fig:Cpsiwcriticalzerowinding}  The zero-flux sector data for the columnar order parameter correlation function at criticality on the honeycomb lattice decays as a power-law, with an exponent whose value is estimated to be $\eta_{\psi}(w_c) \approx 1.72(2)$. Note that this power-law exponent matches the universal value (within error) our earlier unrestricted estimate shown in Fig.~\ref{fig:criticalcolumnarpowerlaw}.}
\end{figure}

In Fig.~\ref{fig:Cpsizerowinding}, we display the zero-flux sector data for the correlation function of the three-sublattice (columnar) order parameter in the honeycomb lattice case. From the quality of the power-law fits on the short-loop side of the transition, it is clear that the short-loop phase has power-law three-sublattice (columnar) order. The value of the stiffness $g$ extracted from these power-law exponents also matches within errors our earlier unrestricted estimates at the same values of $w$. In addition, we see that the correlation function decays rapidly, faster than any power law, in the long-loop phase.
At criticality, we also find that the zero-flux sector result for this correlation function fits well to a power-law form with a critical exponent that is consistent within errors with our earlier unrestricted estimate. This is shown in Fig.~\ref{fig:Cpsiwcriticalzerowinding}.

\section{Discussion}
\label{sec:Discussion}
The flux-fractionalization transition at $w=w_c$ represents an unusual mechanism for the destruction of power-law columnar order. Although topological in nature, it is clearly very different from the well-studied Kosterlitz-Thouless mechanism that involves the proliferation of vortices that invalidate the Gaussian ``spin-wave'' approximation to two-dimensional systems with $U(1)$ symmetry.  In the dimer-loop model, vortices in the height field are explicitly forbidden on either side of the transition. 
Nevertheless, we have already seen that the properties at $w=w_c$ cannot be explained in terms of the Gaussian height action that provides a valid description of the two phases on either side of this critical point. 

Since the exponent $\nu$ that characterizes this transition is, within our numerical errors, equal to the Ising value of $\nu = 1$, it is natural to ask: Is this unusual flux-fractionalization transition in the Ising universality class, with some ``hidden'' Ising order parameter that is not directly related in a simple way to the dimer and loop (equivalently, kagome or planar pyrochlore spin) variables? 
In this scenario, the physical quantities of interest to us here would have to correspond to some geometric degrees of freedom which are not the usual objects of study at an Ising transition.
 
Also, given that the low-temperature physics of such  anisotropic spin $S=1$ systems on the kagome lattice is predicted to have these striking features, the question of possible experimental realizations also becomes interesting.  Below, we provide a brief discussion of these questions.

\subsection{Ising transition?}
\begin{figure} 
	\includegraphics[width=0.49\linewidth]{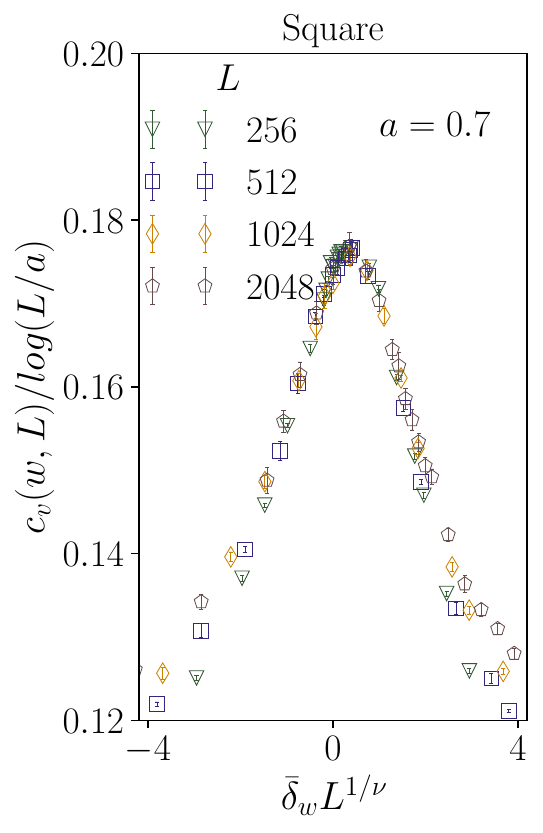}
	\includegraphics[width=0.49\linewidth]{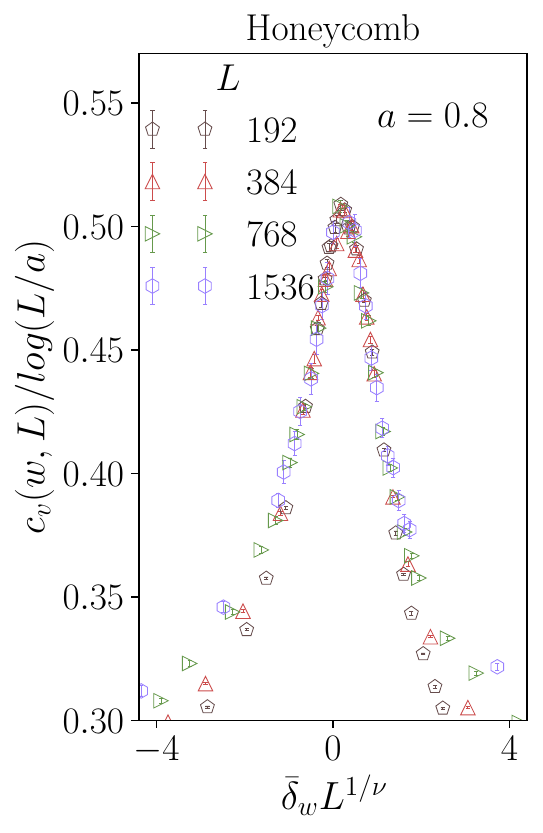}
	\caption{\label{fig:specificheat}  The singular behavior of the specific heat in the vicinity of the flux fractionalization transition on both the square and the honeycomb lattice appears to be logarithmic in nature, as evidenced by the reasonably good scaling collapse exhibited by our data in the vicinity of the critical point. However, we caution that it is impossible to use this dataset to rule out a small power-law divergence.}
\end{figure}

Leaving aside the question of identifying the ``hidden'' Ising order parameter that would be central to this scenario, we focus instead a simple test by studying the critical behavior of the specific heat in the vicinity of the transition. Apart from prefactors that do not affect this critical behavior, the specific heat $c_v$ is proportional to $\frac{1}{L^2}\langle (\delta n_d)^2 \rangle $, where $\delta n_d = n_d - \langle n_d \rangle$ measures the fluctuations in $n_d$, the total number of dimers in a dimer-loop configuration (in equivalent spin language, the total number of spins that have $z$ polarization $S^z = -1$). With this in mind, we define the dimensionless specific heat as 
\begin{eqnarray}
c_v & = & \frac{1}{L^2}\langle (\delta n_d)^2 \rangle  \; .
\end{eqnarray}
In Fig.~\ref{fig:specificheat}, we display our data for $c_v$ in the vicinity of the flux fractionalization transitions on the square and the honeycomb lattice. We see that  the data is consistent with a logarithmic singularity at $w_c$, which would be the expected behavior at an Ising critical point in two dimensions~\cite{McCoy_Wu_1973}. However, we caution that we are unable to rule out a power-law divergence with a small exponent $\alpha \approx 0.2$. 

Nevertheless, this provides additional motivation to explore the possibility that the unusual flux fractionalization transition identified here corresponds to the onset of ``hidden'' Ising order. In this context, it is interesting to note the analogy with generalized $xy$ models which have both $\pi$-periodic and $2\pi$-periodic couplings. Such models have half-vortices in addition to the usual integer vortices. In a narrow range of the phase diagram, such generalized $xy$ models are known to have Ising-like transitions between a high-temperature disordered phase and a low-temperature quasi-long range ordered phase~\cite{Shi_Lamacraft_Fendley_2011,Serna_Chalker_Fendley_2017}. We caution however that this analogy to such Ising transitions is very far from perfect, since the flux fractionalization studied here occurs in a situation characterized by the complete absence of all vortices.

A related question, whose answer may be instructive, has to do with the value of $w_c$: Recall that the measured values of $w_c$ on the square and the honeycomb lattice are both extremely close to $z/2$, where $z$ is the coordination number of the lattice. Indeed, on the square lattice, our error bars do not allow us to rule out the possibility that $w_c = z/2 \equiv 2$ is an exact statement. Why does $w_c$ correspond so closely to $z/2$ on both lattices? Does this signal the possibility of an exact solution on the square lattice, or, at the very least, an accurate approximation scheme for the phase diagram on both lattices?

\subsection{Comments on potential experimental realizations}

From the point of view of potential experimental realizations of the Hamiltonian of Eq.~\ref{eq:frustratedmagnet}, the kagome geometry renders at least one feature quite natural, since the effects of spin-orbit coupling in a two-dimensional kagome layer in an insulating quasi-two dimensional magnet can indeed single out the common $z$ axis perpendicular to the kagome layer.  In such an experimental realization, the $S=1$ moments of Eq.~\ref{eq:frustratedmagnet} are of course expected to the net moments of spin-orbit coupled multiplets in this geometry. 

The main subtlety in identifying possible experimental realizations however lies elsewhere, in the fact that the physics discussed here is a consequence of the competition between a large exchange anisotropy $J_z \gg J_{\perp}$, and a ${\mathcal O}(J_z)$ single-ion anisotropy $\Delta$ that favors the $S^z = 0$ spin polarization for each spin $S=1$ moment. This is quite different from situations in which a large single-ion anisotropy of the opposite sign, {\em i.e.} a single-ion easy-axis anisotropy, {\em induces} correspondingly strong anisotropy in the exchange couplings. Nevertheless,  within a strong-coupling framework of the type~\cite{Ho_Bhattacharjee_Kim_2013,Rau_Lee_Kee_2014} used earlier in other contexts (for pyrochlore and honeycomb iridates, there does not seem to be any fundamental reason why the interesting regime identified here cannot be realized in some materials with strong  spin-orbit scattering effects.

Another important point in this connection is that the local ``$z$'' axes that Eq.~\ref{eq:frustratedmagnet} refers to need not all be exactly identical. They could in principle refer to a local preferred axis that is slightly different for the three basis sites in a kagome unit cell due to the actual geometry of the kagome layer and the shape of the relevant orbitals. Naturally, this would make it impossible to have a perfectly uniform magnetic field oriented along the local  ``$z$'' direction at each site, since a uniform laboratory field perpendicular to the kagome layer would translate to slightly different magnetic field components along the local ``$z$'' axis for each kagome site. However, and this is key, the physics of the magnetization plateau being discussed here is {\em insensitive } to small variations of the component of the field along the local ``$z$'' axis, as well as the presence of small field components transverse to this local ``$z$'' axis. Indeed, the former does not change the low temperature physics as long as one is on the magnetization plateau, and the latter would induce quantum dynamics that only becomes relevant at correspondingly low temperatures, leaving a large intermediate temperature window in which the classical analysis presented here will remain valid.  A full-fledged analysis of this type is outside the scope of the present work, but constitutes an interesting avenue for follow-up work motivated by our results.

In contrast to this, the planar pyrochlore case is of course less likely to have any direct relevance to experimental systems since the nearest-neighbour connectivity of the planar-pyrochlore lattice is somewhat artificial as far as insulating magnets are concerned. However, recent progress in the design and control of Rydberg atom systems has enabled the formulation and analysis of several proposals for realizing interesting states of quantum matter  ~\cite{Ebadi_etalNature2021,Verresen_Lukin_Vishwanath_2021,Verresen_Vishwanath_2022,Samajdar_Joshi_Teng_Sachdev_2023,Yan_Wang_Samajdar_Sachdev_Meng}.  It would be interesting to explore similar realizations of both the kagome and planar pyrochlore Hamiltonians studied here.

\section{Outlook}
\label{sec:Outlook}

The results reported here open up several possiblities for follow up work.
For instance, it would be interesting to explore such dimer-loop systems in three dimensions and explore their potential connection to the low temperature behavior of three-dimensional frustrated magnets.
It would also be interesting to study generalizations where loops of length $s>2$ have a fugacity $n \neq 1$, and explore the phase diagram in the $(w,n)$ plane, in addition to considering the effect of interactions between loop segments and dimers. In addition to these relatively immediate extensions, there are two other questions that appear interesting. Below, we provide a brief discussion of these two questions along with the additional background needed to motivate and formulate them.

\subsection{Rokhsar-Kivelson point of the quantum dimer-loop model}
The fully-packed non-interacting dimer model also lends itself to an interesting interpretation in terms of the ground state wavefunction of an interacting quantum dimer model on the same lattice, with a fine-tuned form of the interaction between dimers; this is the so-called Rokhsar-Kivelson point in the phase diagram of such quantum dimer models~\cite{Rokhsar_Kivelson_1988,Henley_2004,Castelnovo_Chamon_Mudry_Pujol_2005}. Does this connection generalize in a natural way to the fully-packed dimer-loop model studied here?
The answer turns out to be in the affirmative: As we now show, the dimer-loop model encodes the ground state properties of a class of Rokhsar-Kivelson type Hamiltonians for the singlet sector dynamics of frustrated SU(2) symmetric spin $S=1$ antiferromagnets on the square and honeycomb lattices. 

To establish this, we first note the dimer-loop model partition function $Z(w)$ can be viewed as the square of the norm of the following quantum dimer-loop model wavefunction
\begin{eqnarray}
|\psi(w) \rangle &=& \sum_{{\mathcal C}} w^{n_d({\mathcal C})/2}| {\mathcal C}\rangle \; ,
\end{eqnarray}
where the sum is over all fully-packed dimer-loop configurations.
Now, $|\psi(w) \rangle$ is readily seen to be the ground state of a family of simple and natural Rokhsar-Kivelson (RK) type~\cite{Rokhsar_Kivelson_1988,Henley_2004,Castelnovo_Chamon_Mudry_Pujol_2005} Hamiltonians ${\mathcal H}$ for fully-packed dimers and loops on both lattices. Since this is best explained by an explicit example, we construct such a class of RK Hamiltonians here for the square lattice, restricting ourselves to parameter choices that respect the full symmetry of the square lattice.

 The basic idea is to identify a minimal set of local moves which suffice to go from any dimer-loop configuration to any other in a given fixed flux sector. Corresponding to each such local move, one writes down a kinetic energy term and a potential energy term, with the coefficients of these terms chosen to ensure that the Hamiltonian is a positive operator, and the wavefunction $\psi$ is an exact zero enegy ground state of this Hamiltonian.

While we do not have a rigorous proof of this fact, we believe that one such minimal set for the square lattice dimer-loop model comprises four different types of updates acting on an elementary plaquette that have configurations amenable to the execution of these moves. These updates are: ring exchange moves for parallel dimers on a plaquette, ring exchange moves for parallel loop segments on a plaquette, conversion of a length $s=4$ loop on a plaquette to a pair of parallel dimers on the same plaquette and its inverse, and a move that ``absorbs'' a dimer into a loop that has a segment parallel and adjacent to this dimer and its inverse. 

Corresponding to each such update, we have a kinetic energy term in the RK Hamiltonian, with an independent negative coefficient for each term. Here, we impose  additional restriction that the values of various coefficients are constrained by the symmetries of the square lattice, but this is not the most general choice. This leads us to construct:
\begin{widetext}
	\begin{eqnarray}
		{\mathcal H} &=& -\sum_{p} \left[ \gamma \left(|\leftopen_p\rangle \langle \vld_p|+|\bottomopen_p\rangle \langle \hld_p|+|\rightopen_p\rangle \langle \vdl_p|+|\topopen_p\rangle \langle \hdl_p| \right)  +\alpha \left( |\sql_p \rangle \langle \hdd_p| + | \sql_p \rangle \langle \vdd_p| \right) + \beta|\vll_p\rangle \langle \hll_p| +  |\vdd_p\rangle \langle \hdd_p|+ h.c. \right] +\nonumber \\ 
		&&+\sum_p \left[\sqrt{w}\gamma \left( |\leftopen_p\rangle \langle \leftopen_p|+ |\bottomopen_p\rangle \langle \bottomopen_p|+|\rightopen_p\rangle \langle \rightopen_p|+ |\topopen_p\rangle \langle \topopen_p| \right )+\frac{\gamma}{\sqrt{w}} \left ( |\vld_p\rangle \langle \vld_p|+ |\hld_p\rangle \langle \hld_p|+ |\vdl_p\rangle \langle \vdl_p|+ |\hdl_p\rangle \langle \hdl_p| \right ) \right]  + \nonumber \\
		&&+ \sum_p \left[2\alpha w |\sql_p \rangle \langle\sql_p|+ \frac{\alpha}{w} \left (|\hdd_p\rangle \langle \hdd_p|  +|\vdd_p\rangle \langle \vdd_p| \right ) + \beta \left (|\hll_p\rangle \langle \hll_p| + |\vll_p\rangle \langle \vll_p| \right ) + |\hdd_p\rangle \langle \hdd_p| + |\vdd_p\rangle \langle \vdd_p| \right] 
	\end{eqnarray}
\end{widetext}
In the above, the sums are over elementary plaquettes $p$, the local configuration at $p$ is depicted using the notation already defined in Fig.~\ref{fig:linkobjects}, and $\alpha$, $\beta$, $\gamma$ are all positive and measured in units of the amplitude for the ring-exchange of dimers, which has therefore been set to $1$.
Our results predict that this Rokhsar-Kivelson type Hamiltonian ${\mathcal H}$ has a a $T=0$ flux fractionalization transition at $w_c$. It would therefore be interesting to ask if some version of such a transition survives deviations from this fine-tuned Rokhsar-Kivelson form and exists in the phase diagram of more generic quantum dimer-loop models.

\subsection{Nonzero vortex fugacities}
It is also interesting to explore the effects of nonzero fugacities for vortices. For instance, we can introduce a small fugacity $f_{\frac{1}{2}}>0$ for half-vortices, that is, allow sites that are only touched by one segment of a nontrivial loop instead of by two such segments or by a dimer, so that open strings of length $s > 1$ can arise. In addition, we can also allow unit-vortices with a small fugacity $f_1>0$, that is, allow sites that are not touched by any loop segment or dimer. 

Since the unit-vortex correlator decays as $\sim 1/r^{\eta_v^{(1)}}$  (with $\eta_v^{(1)} = g$ ($\eta_v^{(1)} = \sqrt{3}g/2$) for the square (honeycomb) lattice) throughout the $w>w_c$ short-loop phase, we expect $f_1$ to be a relevant perturbation of the Gaussian effective action in short-loop phase,  since it is clear from the measured $g(w)$ curve (Fig~\ref{fig:gvsw}) that $\eta_v^{(1)} < 4$ throughout this phase. However, since the half-vortex correlator is short ranged for all $w>w_c$, $f_{\frac{1}{2}}$ is expected to be an {\em irrelevant} perturbation of the Gaussian effective action in this regime. 

This argument suggests that power-law columnar order would survive for small nonzero $f_{\frac{1}{2}}$ when $w>w_c$ as long as $f_1$ is constrained to be zero, {\em i.e.} power-law columnar order would survive in the presence of a small nonzero density of open strings when $w>w_c$, as long as there are no sites that are left completely untouched by any loop segment or dimer. This is an extremely interesting possibility, since we have already argued that the transition that marks the high-field termination of the one-third magnetization (half magnetization) plateau in kagome (planar pyrochlore) magnets is described precisely by a dimer-loop model with $f_{\frac{1}{2}} \neq 0$ but $f_1 = 0$.

By a very similar argument, we also expect  $f_{\frac{1}{2}}$ to be a relevant perturbation of the Gaussian effective theory for $w<w_c$, since the half-vortex correlator decays as a power-law with exponent $\eta_v^{(1/2)}$ (with $\eta_v^{(1/2)} = g/4$ ($\eta_v^{(1/2)} = \sqrt{3}g/8$) for the square (honeycomb) lattice)), and it is clear from the measured $g(w)$ curves (Fig~\ref{fig:gvsw}) that $\eta_v^{(1/2)} < 4$ throughout this flux-fractionalized phase. What about $f_1$ in this regime? The unit-vortex correlator measured in our unit-vortex worm update is short-ranged throughout this phase. If we take this at face value, the conclusion is that $f_1$ is irrelevant for $w<w_c$. By this argument, if $f_{\frac{1}{2}}=0$ but $f_1$ takes on a small nonzero value, {\em i.e.} if we allow a small density of sites which are not touched by any loop segment or dimer, but do not allow any open strings of length $s>1$, the long-distance properties of the dimer-loop model should continue to admit a description in terms of the Gaussian effective theory for $w<w_c$, but with a renormalized value of $g$. 

Both of these are rather surprising conclusions, for which we do not have any detailed microscopic justifications that go beyond the simple-minded scaling ideas described here. A numerical study aimed at checking the validity of these arguments would thus be of considerable theoretical interest as well.

{\em Acknowledgments:} We thank F.~Alet, S.~Bhattacharjee, D.~Dhar, A.~Gadde, G.~Mandal, S.~Minwalla, and G.~Sreejith for useful discussions.   We are also grateful to N. Bultinck for pointing out to us the potentially interesting analogy between Ising-like flux-fractionalization transitions studied here and  superfluid-insulator transitions in the Ising universality class~\cite{Shi_Lamacraft_Fendley_2011,Serna_Chalker_Fendley_2017}. We gratefully acknowledge generous allocation of computing resources by the Department of Theoretical Physics (DTP) of the Tata Institute of Fundamental Research (TIFR), and related technical assistance from K. Ghadiali and A. Salve. One of us (KD) would like to thank the Indian Institute for Science Education and Research, Pune for hospitality while the intial draft of this work was being written up, and, separately, during the preparation of a revised and expanded manuscript for resubmission. SK was supported at the TIFR by a graduate fellowship from DAE, India. KD was supported at the TIFR by DAE, India, and in part by a J.C. Bose Fellowship (JCB/2020/000047) of SERB, DST India, and by
the Infosys-Chandrasekharan Random Geometry Center
(TIFR).

\bibliography{references}

\end{document}